\documentclass[preprint,12pt]{elsarticle}

\usepackage{amssymb}
\usepackage{amsmath}
\usepackage{amsthm}
\usepackage{graphics}
\usepackage{hyperref}
\usepackage{color}

\journal{Journal of Computational Physics}

\begin{document}

\begin{frontmatter} 



\title{A Conservative Scheme for Vlasov Poisson Landau Modeling Collisional Plasmas }


\author[label0]{Chenglong Zhang\corref{cor1}\fnref{label1}}
\cortext[cor1]{Corresponding author}
\ead{chenglongzhng@gmail.com}

\author[label0]{Irene M. Gamba\fnref{label2}}

\address[label0]{The University of Texas at Austin, Austin, TX 78712, USA}

\fntext[label1]{The Institute for Computational Engineering and Sciences (ICES), The University of Texas at Austin,
201 E. 24th Street, Austin, TX 78712, USA, chenglongzhng@gmail.com }
\fntext[label2]{Department of Mathematics \& ICES, The University of Texas at Austin,
201 E. 24th Street, Austin, TX 78712, USA, gamba@math.utexas.edu}

\begin{abstract}
We have developed a deterministic conservative solver for the inhomogeneous Fokker-Planck-Landau equation coupled with the Poisson equation, which is a  {classical mean-field} primary model for collisional plasmas. Two subproblems, i.e. the Vlasov-Poisson problem and homogeneous Landau problem, are obtained through time-splitting methods, and treated separately by the Runge-Kutta Discontinuous Galerkin method and a conservative spectral method, respectively. To ensure conservation when projecting between the two different computing grids, a special conservation routine is designed to link the solutions of these two subproblems.  {This conservation routine accurately enforces conservation of moments in Fourier space.} The entire numerical scheme is implemented with parallelization with hybrid MPI and OpenMP. Numerical experiments are provided to study linear and nonlinear Landau Damping problems and two-stream flow problem as well. 

\end{abstract}


\end{frontmatter}


\section{Introduction}
\label{sec:intro}

The plasma dynamics is governed by infinite-range interactions, i.e. Coulomb potentials, and thus behaves differently  {from} ordinary molecular gases.
At the kinetic level, among various plasma models, the \emph{Vlasov-Poisson} (VP) equations and \emph{Fokker-Planck-Landau} (FPL) equations are the most representative ones describing, respectively, collisionless and collisional plasma systems.

The VP system is a nonlinear kinetic system modeling the transport of charged particles
in a collisionless plasma, under the effect of a self-consistent electrostatic field and possibly
an externally supplied field. The electrostatic potential is coupled through the Poisson equation. Some natural plasma, as for example solar wind, behaves as
collisionless, since the mean free path of a particle traveling from the Sun to the Earth is of the order of Sun-Earth distance.
Because of its comparative simplicity, numerical schemes for VP equations have been not only thoroughly explored but also well developed. One can refer to, for
example \cite{CGM_RKDG_VP,  CG_VP_infhomostellar, HGMM_DGVP}. The collisionless VP system exhibits a variety of dynamical phenomena. For example,
the well-known filamentation (filaments in phase space and steep gradients in $v$) due to its dispersive nature and Landau damping mechanism for near equilibrium states satisfying some conditions. Readers can refer to \cite{Chen_PlasmaPhysics} for more physical insights.

If collisions are taken into account, particles are scattered and things could be different. To our best knowledge, there is rare work on such models. Thus, we expect to
study the numerical behaviors of the inhomogeneous FPL system for multiple species. The transport of probability density for the particle species $\alpha$ is given by
\begin{equation}\label{InhomogFPL}
    \partial_{t} f_{\alpha} + v \cdot\nabla_{x} f_{\alpha} + F(t,x)\cdot\nabla_{v} f_{\alpha} = \sum_{\beta}a_{\alpha\beta}Q_{\alpha,\beta}(f_{\alpha},f_{\beta}), \qquad v \in \mathbb{R}^{d_{v}}, x \in \Omega_{x} \subseteq \mathbb{R}^{d_{x}} \, ,
\end{equation}
subject to some initial and boundary conditions on $f_{\alpha}$. Here, $f_{\alpha}$ is the distribution for species $\alpha$, the term $Q_{\alpha,\beta}(f_{\alpha},f_{\beta})$ is a nonlinear, nonlocal operator in divergence form and models the $(\alpha,\beta)$ pair collisions (e.g. electron-electron, ion-ion, electron-ion, etc.) and $a_{\alpha\beta}$ are the coupling parameters. In our present work, we take $a_{\alpha\beta}=\frac{1}{\varepsilon}$ to be the collision frequency with $\varepsilon$ the Knudsen number. The case $a_{\alpha\beta} \rightarrow 0$ corresponds to the Vlasov-Poisson system. The force field $F(t,x)$ only depends on time and space position and can be external or self-consistent.
If it is self-consistent, it corresponds to the electrostatic force $qE(t,x)$, where $q$ is the charge and $E(t,x)$ is the self-consistent electrostatic field obtained from the Poisson equation for charges
\begin{equation}\label{Poisson}
    E(t,x)=-\nabla_{x} \Phi(t,x); \qquad -\Delta_{x}\Phi=\sum_{\beta}\int_{\mathbb{R}^{3}}f_{\beta}(v) dv \, ,
\end{equation}
subject to some boundary condition on $\Phi$.

The FPL transport equation is used to model long-range Coulomb interactions between charged particles (e.g binary collisions occurring in a plasma). It is of primary importance in modeling
evolution of collisional plasma and actually a rather realistic model especially when the magnetic field is very weak. The FPL transport equation can be derived from the general Boltzmann transport equation by taking the so-called binary grazing collision limit, i.e collisions that only result in very small
deflections of particle trajectories, as is the case for Coulomb potentials with Rutherford scattering \cite{Rutherford}. The original derivation is due to Landau \cite{Classical_Landau}. It is also often called Fokker-Planck-Landau equation in Plasma Physics due to the independent derivation in the Fokker-Planck form in \cite{FP_Landau}.

With the general non-isotropic Landau collision operator $Q$, the inhomogeneous FPL model gains huge difficulties to handle, both analytically and numerically. The main factors generating such
difficulties are the nonlinearity, non-locality and diffusive nature with high dimensionality. Unlike other kinetic models, for example Boltzmann equations, where some non-deterministic methods (DSMC) have been successfully applied,
the infinite-range potential interactions greatly limit the applications of these type of  Monte Carlo methods. Many  have tried to develop efficient deterministic solvers for the inhomogeneous FPL equations. However,
due to the computational complexity mentioned above, they have turned to some simplified versions of this problem. Among them, the space homogeneous Landau equations in the isotropic case were study in  \cite{Buet_homoFPL_iso}, the 1D Fokker-Planck type operator \cite{Eulerian_1D_FPL, Eulerian_1D_FPL_Erratum}, the cylindrically symmetric problem in \cite{Lemou_cylindricaFPL}, as well as very recent work in  \cite{TaitanoChacon2015,TaitanoChacon2016} on a conservative scheme for a multispecies system of FPL equations. 
 
L. Pareschi et al. proposed a spectral method to solve FPL equations \cite{fastSpectralFPL}, by taking truncated Fourier series and extending solutions by periodicity. This method was not
intended to preserve the natural collision invariants, so, as a consequence they  introduced unphysical binary collisions. It cannot avoid \emph{aliasing effects}, which will be present whenever a vanishing function is approximated by a periodic
one. Later, Filbet and Pareschi \cite{FPL_1x2v} applied the spectral method to study inhomogeneous FPL with 1D in space and 2D in velocity. The pure transport equation was further split and a finite volume scheme
was used. Then, Crouseilles and Filbet \cite{FPL_1x3v} proposed a solver for inhomogeneous FPL with 1D in space and 3D in velocity, where the pure transport part was treated with a finite volume scheme and the Landau operator
was approximated by averaging of uncentered finite difference operators. However, the solver in \cite{FPL_1x3v} only preserved mass and energy at the discrete level (for the uncentered finite difference approximate Landau operator), under some symmetry assumptions on the initial datum.

 {At the time of writing this manuscript, we were introduced to the work by Dimarco et al.\cite{Dimarco_colPlasma}. Here, we find it necessary to briefly compare it with our work. They followed the scheme of time splitting, using a semi-Lagrangian method for the collisionless part and the spectral method for the collisional part. An Asymptotic-Preserving (AP) strategy is also applied to handle the stiffness due to the small Knudsen number. We should point out, we are focusing on rather different perspectives. Our goal is not approximating the fluid limit but aiming at weak to moderately strong collisions. In addition to the shortcomings similar to \cite{fastSpectralFPL} mentioned above, the solver in \cite{Dimarco_colPlasma}, especially the AP scheme, takes advantage of the assumption that the states should be close to Gaussian. However, our target problems allow states to be far from equilibrium. We maintain conservation during the entire life of simulation. This is achieved by a novel routine that ensures no conservation will be damaged when projecting between Fourier and DG spaces. The conservation routine, as shown by the second author \cite{AGT_convSpcBoltzmann}, is crucial for the evolution of the probability distribution to a Gaussian. The DG solver for the Vlasov-Poisson subproblem is readily to approximate functions that are less ``smooth" and easily incorporate more non-standard boundary conditions. We tested our scheme in 3D velocity space while they only tested in 2D. We actually applied a much coarser mesh grid in spectral as much as in DG space but still achieved numerical results that agree quite well with theoretical benchmark. In addition, we conducted extensive comparisons with theoretical benchmarks, esp. including the electron-ion system which exhibits the necessity of conservation properties. At last, our implementations are all done in parallel with HPC techniques.}

In this work, we follow a  {standard} time-splitting scheme, splitting the original inhomogeneous FPL equation into a pure transport problem ( i.e  Vlasov-Poisson equation for advection ) and a homogeneous FPL equation for collisions. These two subproblems can be treated with completely different schemes. For the VP problem, we apply the RKDG method with a piecewise polynomial basis subspace covering all collision invariants, which can be proved to conserve mass, momentum and kinetic energy up to some boundary error terms that disappear if the domain is taken large enough.
While for the homogeneous FPL equation, different than in \cite{fastSpectralFPL}, we extend the spectral method first introduced in \cite{GT_jcp} for the nonlinear Boltzmann transport equation and propose a conservative spectral method for the homogeneous FPL equation, by first extending the solution by zero, representing the collision integral through choosing Fourier modes as the test functions in the weak form and enforcing conservation routines.
Since two completely different numerical scheme are applied separately, our challenge is not only to link two different meshes and at the same time, but also to keep the conserved quantities. We have designed a new conservation correction process such that, after projecting the
conservative spectral solution onto the DG mesh, the conserved moments are transferred to the DG solution as well. 

This work is based on a section of Thesis Dissertation \cite{Cl_Zhang_thesis2014} of the first author under the direction of the second author of this manuscript.

\section{The Fokker-Planck-Landau Operator}\label{sec:FPL}

The FPL operator models binary collisions in a system of single- or multi-species and reads
\begin{equation}\label{Q_FPL_multi}
    Q_{\alpha,\beta}(f_{\alpha},f_{\beta})=\nabla_{v}\cdot \int_{\mathbb{R}^{3}} \mathbf{S}(v-v_{*})(f_{\beta}(v_{*})\nabla_{v}f_{\alpha}(v)-f_{\alpha}(v)\nabla_{v_{*}}f_{\beta}(v_{*})) dv_{*} \, ,
\end{equation}
with the $d\times d$ nonnegative and symmetric projection matrix
\begin{equation}\label{ProjMatrix}
\mathbf{S}(u)=L|u|^{\gamma+2}(\mathbf{Id}-\frac{u\otimes u}{|u|^{2}})\, ,
\end{equation}
where $\mathbf{Id}$ is the $d\times d$ identity matrix; $\Pi(u)=\mathbf{Id}-\frac{u\otimes u}{|u|^{2}}$ is the orthogonal projection upon the space orthogonal to $u$. It's semi-positive definite with eigenvalues 0,1,1. The constant $L$ is positive(a value related to the logarithm of the dimensionless Debye radius of screening of the Coulomb potential in plasma). For simplicity, we take $L=1$ in the following.

The inverse-power laws has $\gamma\geq -3$. Similar to Boltzmann equations, different $\gamma$ categorizes hard potentials for $\gamma >0$, Maxwellian molecules for $\gamma=0$ and soft potentials for $\gamma <0$. Here, however, we only focus on most interesting case $\gamma=-3$, corresponding to Coulomb interactions.

When $\alpha=\beta$, the operator $Q_{\alpha,\alpha}$ will be a nonlinear (bilinear) integro-differential operator in divergence form. Here and in the following, when talking about single-species distributions, we will drop the subscript $\alpha$ for simplicity. The strong form of this nonlinear partial integrodifferential equation is
\begin{equation}\label{InhomogFPL}
    \partial_{t} f + v \cdot\nabla_{x} f + F(t,x)\cdot\nabla_{v} f = \nu Q(f,f), \qquad v \in \mathbb{R}^{3}, x \in \Omega_{x} \subseteq \mathbb{R}^{3} \, ,
\end{equation}
where the collision kernel is of the form
\begin{equation}\label{Q_FPL_single}
    Q(f,f)=\nabla_{v}\cdot \int_{\mathbb{R}^{3}} \textbf{S}(v-v_{*})(f(v_{*})\nabla_{v}f(v)-f(v)\nabla_{v_{*}}f(v_{*})) dv_{*} \, ,
\end{equation}
and the collision frequency $\nu =\frac{1}{\varepsilon}$ with $\varepsilon$ being the Knudsen number. The case $\nu \rightarrow 0$ corresponds to the collisionless Vlasov-Poisson system.

{The FPL collision operator can also be viewed as as the divergence of a non-local, binary gradient operator $L_F(f,g)$, referred to here as the Landau flux, defined by
\begin{align}\label{landau-flux} 
 Q(f,g)&=\nabla_v \cdot L_{F}(f,g) \, , \nonumber\\
 L_{F}(f,g) &=  \int_{\mathbb{R}^{3}} \textbf{S}(v-v_{*})(f(v_{*})\nabla_{v}g(v)-f(v)\nabla_{v_{*}}g(v_{*})) dv_{*} . 
\end{align}
}

The FPL operator, as a limit of the Boltzmann collision operator, possesses similar conservation laws and decay of entropy($H$-theorem). That is
\begin{equation}
    \int_{\mathbb{R}^{3}} Q(f,f)(v)\phi(v)dv = 0 \, ,
\end{equation}
if and only if
\begin{equation}
\phi(v)=1,v,|v|^{2}
\end{equation}
corresponding to the conservation of mass (charge), momentum and kinetic energy. We call the $d+2$ test functions $\phi(v)=1, v, |v|^{2}$ \emph{collision invariants}.

In addition, for any $f(v)>0$, if we set $\phi(v)=\log f(v)$, one can show the following dissipation of entropy
\begin{equation}
\label{entropy_disp}
\frac{d}{dt}\int_{\mathbb{R}^{d}} f\log f dv=\int_{\mathbb{R}^{d}}Q(f,f)(v)\log f(v)dv \leq 0 \, ,
\end{equation}
which also implies the equilibrium states given by the Maxwellian distribution
\begin{equation}\label{Maxwellian}
    M(x,v)=\frac{\rho}{(2\pi k_{B}T)^{\frac{3}{2}}}\exp\left(-\frac{|v-\bar{v}|^{2}}{2k_{B}T}\right) \, ,
\end{equation}
where $k_{B}$ is the Boltzmann constant. The local dependence of $x$ is from the mass $\rho(x)$, the mean velocity $\bar{v}(x)$ and the kinetic temperature $T(x)$, given by
\begin{equation}\label{MacroQuantitiy}
    \rho=\int_{\mathbb{R}^{3}}f(x,v) dv, \qquad \bar{v}=\frac{\int_{\mathbb{R}^{3}}f(x,v)v dv}{\rho}, \qquad T = \frac{\int_{\mathbb{R}^{3}}f(x,v)|v-\bar{v}|^{2} dv}{3\rho} \, .
\end{equation}

When $\alpha\neq\beta$, the operator $Q_{\alpha,\beta}$ models collisions between two different species. It is essentially a linear operator and
the treatment will be similar and sometimes even much simpler compared with the fully nonlinear one (\ref{Q_FPL_single}). We will consider different problems associated to different forms of the operator $Q_{\alpha,\beta}$ in the following sections.

\section{Time Splitting}\label{sec:tSplit}

The main challenges come from the high dimensionality, nonlinearity, diffusive nature, conservation properties, positivity, etc, which require very careful design of the numerical scheme. We divide and conquer starting from a time splitting method.
For zero force field, i.e $F(t,x)=0$, the time-splitting is an efficient and reliable way for conquering inhomogeneous problems; however, we will employ the time-splitting to non-zero force field as well and show that it also works.

We discretize time $t_{n}=t_{0}+n\Delta t$, where $\Delta t$ is the time step size. Denote $f_{n}(x,v)=f(t_{n}, x, v)$. In a time interval $[t_{n}, t_{n+1}]$,  a first order time splitting scheme turns the original problem into two subproblems
\begin{eqnarray}\label{eqn:timesplitting_vp}
\text {(1) The Vlasov (Collisionless) Problem} \nonumber \\
\partial_{t} g(x, v, t) + v \cdot \bigtriangledown_{x} g(x, v, t) + F(t,x)\cdot\nabla_{v}g  =  0 \, , \nonumber \\
g(0, x, v) = f_n(x, v) \, ,
\end{eqnarray}
and
\begin{eqnarray}\label{eqn:timesplitting_fpl}
\text{(2) The Homogenous FPL (Collisional) Problem} \nonumber \\
\partial_{t}\tilde{f}(x, v, t) = \frac{1}{\varepsilon}Q(\tilde{f}, \tilde{f}) \, , \nonumber \\
\tilde{f}(0, x, v) = g(\Delta t, x, v ) \, .
\end{eqnarray}

If we denote the above solution operators \eqref{eqn:timesplitting_vp} and \eqref{eqn:timesplitting_fpl} by $A_{n}(\Delta t)$ and $H_{n}(\Delta t)$, respectively. then the solution at time step $t_{n+1}$ is given by
\begin{equation}\label{timesplit}
    f_{n+1}(x,v)=H_{n}(\Delta t)\circ A_{n}(\Delta t)f_{n}(x,v) \, .
\end{equation}

\noindent\textbf{Remark. }
\textit{This splitting is first order in time. Higher order time splitting is also possible. For example, one common scheme is \emph{Strang splitting}, which  {is} second order in time.}

The above two steps can be performed with different methods. The collisionless step can be done with finite difference, finite volume or (DG)FEM; while the collisional step requires special techniques to handle the collisional operator. They will be introduced in the following sections.

\section{The Conservative Spectral Method for Homogeneous FPL Equation}\label{sec:FPL_spectra}

As mentioned in the time splitting scheme above, the collisionless and collisional subproblems can be treated separately with different methods. In the current section, we restrict ourselves to the homogeneous FPL equation for the most interesting Coulombian case, $\gamma=-3$, in 3d velocity space.

Different  {from} the one proposed in \cite{fastSpectralFPL}, by taking truncated Fourier series and extending solutions by periodicity, we don't
have to introduce nonphysical binary collisions and simply extend the solution by zero. Conservation of moments are guaranteed by calling a conservation routine.

\subsection{Domain of Computation}

We assume that the distribution function $f$, the solution of the FPL equation, usually is not compactly supported in $v$ but is of negligible mass outside of a finite ball
\begin{equation*}
    B_{L}(\bar{v})=\{v\in \mathbb{R}^{3}: |v-\bar{v}|\leq R \}\, ,
\end{equation*}
where $\bar{v}$ and $R$ actually depends on $x$ in the inhomogeneous case. However, numerically, in order to find an approximation in a finite domain, we assume $f$ is compactly supported in the above ball.

Consider the cube
\begin{equation*}
    \Omega_{v}=\{v\in \mathbb{R}^{3}: |v_{i}-\bar{v}_{i}|\leq L_{v}, \, i=1,2,3\}\, ,
\end{equation*}
which contains $B_{L}(\bar{v})$. This cube will be defined as the domain of computation for all velocity variables.

For the sake of simplicity, we assume a uniform discretization over the domain and also $\bar{v}=0$. Let $N$ be the number of discretizations in each direction of velocity, then the mesh for each direction of velocities is
\begin{equation}\label{mesh_requirement}
    h_{v}=\frac{2L_{v}}{N}, \qquad v_{i}=-L_{v}+ih_{v}, \, 0\leq i<N \, .
\end{equation}
In order to employ the standard FFT package \cite{fft}, the corresponding mesh for the Fourier space should satisfy
\begin{equation} \label{eqn:FFT_requirement}
    h_{v}h_{\xi}=\frac{2\pi}{N}, \qquad L_{\xi}=\frac{N}{2}h_{\xi}, \qquad \xi_{i}=-L_{\xi}+ih_{\xi}, \quad 0\leq i<N \, ,
\end{equation}
where $h_{v}$ and $L_{v}$ , $h_{\xi}$ and $L_{\xi}$ are the mesh size and cube side-length for the velocity and Fourier domain, respectively.

The whole mesh for the cubic domain will be the tensor product of the mesh on each direction.

\subsection{Spectral Representation}
\label{sec:SpecRep}

We first look at the weak form of the FPL integrals. Suppose $\varphi(v)$ is smooth over the whole domain and the unknown $f$ has exponentially decaying tails when $|v|\rightarrow\infty$ with some rate. For the sake of simplicity, we drop the dependence on variables $t$ and $x$.

Then, the weak form of the FPL operator is
\begin{equation}\label{FPLWeak1}
\begin{split}
 \int_{\mathbb{R}^{3}} Q(f,f)\varphi(v)dv &=-\int_{\mathbb{R}^{3}}\int_{\mathbb{R}^{3}}
\textbf{S}(v-v_{*})(f_{*}\nabla f-(\nabla f)_{*}f)\cdot\nabla_{v}\varphi(v))dv_{*} dv\\
&=\int_{\mathbb{R}^{3}}\int_{\mathbb{R}^{3}}(\nabla_{v_{*}}\varphi(v_{*})-\nabla_{v}\varphi(v))^{T}\textbf{S}(v-v_{*})f_{*}\nabla f dv_{*}dv \, .
\end{split}
\end{equation}

Let $\varphi(v)=(2\pi)^{-3/2}e^{-i\xi\cdot v}$ be the Fourier multiplier, and $u=v-v_{*}$.

Then,
\begin{equation}\label{FPL_FT}
\begin{split}
 \widehat{Q}(\xi)&= \int_{\mathbb{R}^{3}}\int_{\mathbb{R}^{3}}\mathbf{S}_{kl}(v-v_{*})(\partial_{k}\varphi(v_{*})-\partial_{k}\varphi(v))f(v_{*})\partial_{l} f(v) dv_{*}dv \\
&= (2\pi)^{-3/2}\int_{\mathbb{R}^{3}}\int_{\mathbb{R}^{3}}\mathbf{S}_{kl}(v-v_{*})(-i\xi_{k})e^{-i\xi\cdot v}(e^{-i\xi\cdot(v_{*}-v)}-1)f(v_{*})\partial_{l} f(v) dv_{*}dv \\
&=\int_{\mathbb{R}^{3}} du \mathbf{S}_{kl}(u)(-i\xi_{k})(e^{i\xi\cdot u}-1)\left((2\pi)^{-3/2}\int_{\mathbb{R}^{3}} \tau_{u}f(v)\partial_{l}f(v)e^{-i\xi\cdot v}dv \right) \\
&=(2\pi)^{-3/2}\int_{\mathbb{R}^{3}} \widehat{\tau_{u}f}\ast \widehat{\partial_{l}f} (\xi)\mathbf{S}_{kl}(u)(-i\xi_{k})(e^{i\xi\cdot u}-1) du \\
&=\int_{\mathbb{R}^{3}} d\omega \xi_{k}\omega_{l}\widehat{f}(\xi-\omega)\widehat{f}(\omega) \left((2\pi)^{-3/2}\int_{\mathbb{R}^{3}}\mathbf{S}_{kl}(u)(e^{i\omega\cdot u}-e^{-i(\xi-\omega)\cdot u})du\right) \\
&=\xi_{k}\int_{\mathbb{R}^{3}} [\widehat{\mathbf{S}}_{kl}(-\omega)-\widehat{\mathbf{S}}_{kl}(\xi-\omega)]\omega_{l}\widehat{f}(\xi-\omega)\widehat{f}(\omega) d\omega \\
&=\int_{\mathbb{R}^{3}}  \left( \widehat{f}(\xi-\omega)\widehat{f}(\omega)\omega^{T}\widehat{\mathbf{S}}(\omega)\omega - (\xi-\omega)^{T}\widehat{\mathbf{S}}(\omega)(\xi-\omega)\widehat{f}(\xi-\omega)\widehat{f}(\omega)\right) d\omega \, ,
\end{split}
\end{equation}
where there is a summation over the same subscript indices.

Another weak form that is of interest is given by
\begin{equation}\label{FPLWeak2}
\begin{split}
 &\int_{\mathbb{R}^{3}} Q(f,f)\varphi(v)dv  =\int_{\mathbb{R}^{3}}\int_{\mathbb{R}^{3}}(\nabla_{v_{*}}\varphi(v_{*})-\nabla_{v}\varphi(v))^{T}\textbf{S}(v-v_{*})f_{*}\nabla f dv_{*}dv \\
&= \int_{\mathbb{R}^{3}}\int_{\mathbb{R}^{3}} ff_{*} \left( 2[\nabla_{v} \cdot\textbf{S}(v-v_{*})]\cdot\nabla_{v}\varphi(v) +\mathbf{S}(v-v_{*})\colon\nabla^{2}_{v}\varphi(v)\right) dvdv_{*} \, .
\end{split}
\end{equation}
In addition, with the same derivation, we have
\begin{equation}\label{WeightedConv}
\begin{split}
    \widehat{Q}(f,f)(\xi) &=\int_{\mathbb{R}^{3}}\int_{\mathbb{R}^{3}} ff_{*}e^{-i\xi\cdot v}G(\xi,u) dv du \\
    &=\int_{\mathbb{R}^{3}} \widehat{f}(\xi-\omega)\widehat{f}(\omega)\widehat{G}(\xi,\omega) d\omega \, .
\end{split}
\end{equation}
where the precomputed weight in Fourier domain $\widehat{G}(\xi,\omega)$ is the same as given by the above \eqref{FPL_FT}, and the weight in velocity domain is
\begin{equation}\label{SpatialWeight}
    G(\xi, u)=|u|^{-3}\left(i4u\cdot\xi-|u|^{2}|\xi^{\perp}|^{2}\right) \, ,
\end{equation}
where $\xi^{\perp}=\xi-(\frac{\xi \cdot u}{|u|})\frac{u}{|u|}$. We point out that \eqref{SpatialWeight} can be also retrieved from the Fourier transform representation of the Boltzmann collision operator written as a weighted convolution of Fourier transforms. It is recently shown in \cite{HG_grazinglimit} that the weight corresponding to the Boltzmann collision operator converges to the one for Landau operator, if collisions are grazing and the solutions of the BTE have some regularity and decay for large velocity.


It is easy to see that the above weighted convolution \eqref{FPL_FT}, since variables $\omega$ and $\xi-\omega$ are separable in the weights, leads to an $N^{d}\log(N)$ scheme (where $N$ is the number of discretizations on each direction), when FFT is applied. In addition, the weights can be pre-computed and only have to be computed once. We will also derive the above weight analytically, without any extra integral approximations.

Using the same notations to denote the truncated transforms (i.e integrated over some ball $u\in B_{R}(0)$ instead of the whole domain), we write
\begin{equation}\label{S_FT}
\widehat{\mathbf{S}_{kl}}(\omega) =(2\pi)^{-3/2}\int_{B_{R}(0)}{\mathbf{S}}_{kl}(u)e^{-i\omega\cdot u}du \, .
\end{equation}
In addition, they can be decomposed into
\begin{equation}\label{S_FTsplit}
\widehat{\mathbf{S}_{kl}}(\omega) = \widehat{\mathbf{S}^{1}_{kl}}(\omega)-\widehat{\mathbf{S}^{2}_{kl}}(\omega) \, ,
\end{equation}
with
\begin{equation*}
\begin{split}
\widehat{\mathbf{S}^{1}_{kl}}(\omega)&=(2\pi)^{-3/2}\int_{B_{R}(0)}|u|^{\gamma+2}\delta_{kl}e^{-i\omega\cdot u}du \\
\widehat{\mathbf{S}^{2}_{kl}}(\omega)&=(2\pi)^{-3/2}\int_{B_{R}(0)}|u|^{\gamma}u_{k}u_{l}e^{-i\omega\cdot u}du \, .
\end{split}
\end{equation*}

It is not hard to observe the following symmetry properties of $\widehat{\textbf{S}_{kl}}(\omega)$
\begin{equation}\label{Sym_Shat}
\begin{split}
& \widehat{\textbf{S}^{2}_{11}}(\omega_{1},\omega_{2},\omega_{3}) =  \widehat{\textbf{S}^{2}_{33}}(\pm\omega_{2},\pm\omega_{3},\pm\omega_{1})=\widehat{\textbf{S}^{2}_{33}}(\pm\omega_{3},\pm\omega_{2},\pm\omega_{1}) \, ,\\
&\widehat{\textbf{S}^{2}_{22}}(\omega_{1},\omega_{2},\omega_{3}) =  \widehat{\textbf{S}^{2}_{33}}(\pm\omega_{1},\pm\omega_{3},\pm\omega_{2})=\widehat{\textbf{S}^{2}_{33}}(\pm\omega_{3},\pm\omega_{1},\pm\omega_{2}) \, ,
\\
&\widehat{\textbf{S}^{2}_{12}}(\omega_{1},\omega_{2},\omega_{3}) = \widehat{\textbf{S}^{2}_{21}}(\omega_{1},\omega_{2},\omega_{3}) =  \widehat{\textbf{S}^{2}_{13}}(\omega_{1},\pm\omega_{3},\omega_{2})=-\widehat{\textbf{S}^{2}_{13}}(-\omega_{1},\pm\omega_{3},\omega_{2}) \, ,\\
&\widehat{\textbf{S}^{2}_{23}}(\omega_{1},\omega_{2},\omega_{3}) =\widehat{\textbf{S}^{2}_{32}}(\omega_{1},\omega_{2},\omega_{3}) =  \widehat{\textbf{S}^{2}_{13}}(\omega_{2},\omega_{1},\omega_{3}) \, .
\end{split}
\end{equation}
Therefore, we only need to study, say, $\widehat{\textbf{S}^{1}_{11}}$, $\widehat{\textbf{S}^{2}_{33}}$ and $\widehat{\textbf{S}^{2}_{13}}$. See Appendix for detailed derivations.

Then, by considering the symmetry properties \eqref{Sym_Shat}
\begin{equation}\label{}
   \widehat{\textbf{S}}(\omega)=\left(
                 \begin{array}{ccc}
                   \widehat{\textbf{S}^{1}_{11}}(\omega)-\widehat{\textbf{S}^{2}_{33}}(\omega_{2},\omega_{3},\omega_{1}) & -\widehat{\textbf{S}^{2}_{13}}(\omega_{1},\omega_{3},\omega_{2}) &  -\widehat{\textbf{S}^{2}_{13}}(\omega) \\
                     -\widehat{\textbf{S}^{2}_{13}}(\omega_{1},\omega_{3},\omega_{2}) & \widehat{\textbf{S}^{1}_{11}}(\omega)-\widehat{\textbf{S}^{2}_{33}}(\omega_{1},\omega_{3},\omega_{2})  & -\widehat{\textbf{S}^{2}_{13}}(\omega_{2},\omega_{1},\omega_{3})\\
                     -\widehat{\textbf{S}^{2}_{13}}(\omega) & -\widehat{\textbf{S}^{2}_{13}}(\omega_{2},\omega_{1},\omega_{3}) & \widehat{\textbf{S}^{1}_{11}}(\omega)-\widehat{\textbf{S}^{2}_{33}}(\omega) \\
                 \end{array}
               \right)\, ,
\end{equation}
we observe that, if we write $\widehat{\textbf{S}}(\omega)$ as
\begin{equation}
\widehat{\textbf{S}}(\omega)=2\sqrt{\frac{2}{\pi}}\frac{R|\omega|-\sin(R|\omega|)}{R|\omega|^{3}}\widetilde{\Pi}(\omega) \, ,
\end{equation}
$\widetilde{\Pi}(\omega)$ is an orthogonal projection onto $\omega$, i.e. $\widetilde{\Pi}(\omega)\omega=\omega$.
Thus the weighted convolution becomes
\begin{equation}\label{Qhat_analytical}
\begin{split}
\widehat{\textbf{Q}}(\widehat{f},\widehat{f}) &=\int_{\Omega_{\xi}}  \left( \widehat{f}(\xi-\omega)\widehat{f}(\omega)\omega^{T}\widehat{\textbf{S}}(\omega)\omega - (\xi-\omega)^{T}\widehat{\textbf{S}}(\omega)(\xi-\omega)\widehat{f}(\xi-\omega)\widehat{f}(\omega)\right) d\omega \\
&=2\sqrt{\frac{2}{\pi}}\int_{\Omega_{\xi}} \frac{R|\omega|-\sin(R|\omega|)}{R|\omega|} \widehat{f}(\omega)\widehat{f}(\xi-\omega) d\omega  \\
&\quad - \int_{\Omega_{\xi}} (\xi-\omega)^{T}\widehat{\textbf{S}}(\omega)(\xi-\omega)\widehat{f}(\xi-\omega)\widehat{f}(\omega) d\omega \, ,
\end{split}
\end{equation}
where $\Omega_{\xi}=[-L_{\xi},L_{\xi}]^{3}$ with $L_{\xi}$ defined in (\ref{eqn:FFT_requirement}), and the first integral in the above last formula is zero if $|\omega|=0$. It can be readily computed in $O(N^{3}\log(N))$, through FFT.

\subsection{Conservation Routines}
\label{Sec:ConserveRoutine_FPL}
 {The idea of imposing conservation routines have been successfully implemented in conservative spectral or discontinuous Galerkin solvers for the Boltzmann equation, see \cite{GT_jcp, ZhangGamba_DGBE}. Here, we are following a similar argument.}

Let $M=N^{3}$ be the total number of discretizations in the velocity space, i.e the total number of Fourier modes, and
\begin{equation}
\widehat{\textbf{Q}}=\left(\widehat{Q}_{0},\ldots, \widehat{Q}_{M-1} \right)^{T}
\end{equation}
be the vector of Fourier modes in (\ref{Qhat_analytical}), and correspondingly $\textbf{Q}$ be its inverse transform. Denote by
\begin{equation}
\mathbf{F}=\left(F_{0},\ldots, F_{M-1} \right)^{T}
\end{equation}
the distribution vector at current time step. 

 {After having $\widehat{\textbf{Q}}(f,f)(\xi)$, the $Q(f,f)(v)$ will be reconstructed by a partial sum of Fourier series,
\begin{equation}\label{Q_FS}
    Q(f,f)(v)=\frac{(2\pi)^{3/2}}{(2L)^{3}}\sum_{|k| < N^{3}}\widehat{\textbf{Q}}(\xi_{k})e^{i\xi_{k}\cdot v}\, ,
\end{equation}
where $\xi_{k}=\frac{\pi k}{L}$ are the spectral modes, $k=(k_{1},k_{2},k_{3})$ is the multi-index.}

 {Our goal is to find the corrected mode coefficients $\widehat{\textbf{Q}}(\xi_{k})$, such that
\begin{equation}\label{eqn:generalConserve}
    \int_{\Omega_{v}}Q(f,f)(v)\phi(v)dv =0 \, .
\end{equation}
Here, $\phi(v)$ are the collision invariants.}

 {In the literature \cite{GT_jcp}, a conservation routine was designed in velocity space and specially dependent on quadrature, e.g Trapezoidal rule, of evaluating Fourier integrals. We also extended the idea to space homogeneous Landau equations \cite{Cl_Zhang_thesis2014}. However, this sort of conservation routine is sufficient for space homogeneous problem, but not enough for time-splitting scheme for inhomogeneous systems, as will be discussed in Section \ref{sec:linking}.}

 {Here, we propose a new conservation routine independent of any quadrature rule. This is designed in Fourier space rather than velocity space.}

 {Plugging \eqref{Q_FS} back into \eqref{eqn:generalConserve} gives constraints on the corrected mode coefficients. If denote by $\widehat{\mathbf{Q}}_{R}$, $\widehat{\mathbf{Q}}_{I}\in \mathbb{R}^{M}$  the real and imaginary parts of $\widehat{\textbf{Q}}$, respectively, then
\begin{equation}\label{CQ}
    \mathbf{C}_{R}\widehat{\mathbf{Q}}_{R} - \mathbf{C}_{I}\widehat{\mathbf{Q}}_{I}=\mathbf{0}\, ,
\end{equation}
where the constraint matrices $\mathbf{C}_{R}$, $\mathbf{C}_{I}\in \mathbb{R}^{5\times M}$, are the real and imaginary parts of the following
\begin{equation}\label{computeC1}
\mathbf{C}_{R}(l,k)+i\mathbf{C}_{I}(l,k)=\frac{1}{(2L)^{3}}\int_{\Omega_{v}}e^{i\xi_{k}\cdot v}\phi_{l}(v)dv\, ,
\end{equation}
where $\phi_{l}(v)=1,v,|v|^{2}$.}

 {Indeed,
\begin{equation}\label{computeC2}
\begin{split}
    &\mathbf{C}_{R}(0,k)=\prod^{3}_{i=1}\text{sinc}(L\xi_{k_{i}}), \qquad \mathbf{C}_{I}(1,k)=0 \\
    &\mathbf{C}_{R}(l,k)=0, \qquad \mathbf{C}_{I}(l,k)=\begin{cases}
         \frac{\text{sinc}(L\xi_{k_{l}}) - \cos(L\xi_{k_{l}})}{\xi_{k_{l}}}\prod^{3}_{i\neq l}\text{sinc}(L\xi_{k_{i}}) &\text{ $\xi_{k_{l}}\neq 0;$ }\\
         0 & \text{ $\xi_{k_{l}}= 0$}
        \end{cases}, \quad l=1,2,3 \\
     &\mathbf{C}_{R}(4,k)=\sum^{3}_{l=1}\left(\prod^{3}_{i\neq l}\text{sinc}(L\xi_{i})\right)\cdot\begin{cases}
            L^{2}\text{sinc}(L\xi_{l}) - 2\frac{\text{sinc}(L\xi_{l})-\cos(L\xi_{l})}{\xi^{2}_{l}} &\text{ $\xi_{l}\neq 0;$ }\\
         \frac{L^{2}}{3} & \text{ $\xi_{l}= 0$}
        \end{cases}  ,\\
     & \mathbf{C}_{I}(4,k)=0
\end{split}
\end{equation}
}

 {The conservation correction is found by solving the following constrained optimization problem:
Find $\widehat{\mathbf{Q}}=[\widehat{\mathbf{Q}}^{T}_{R}, \widehat{\mathbf{Q}}^{T}_{I}]^{T}\in \mathbb{R}^{2M}$, the minimizer of the optimization problem
\begin{equation}\label{Min}
\begin{split}
& \text{min} \quad \|\widehat{\mathbf{Q}}_{o} - \widehat{\mathbf{Q}}\|^{2}_{2}  \\
& \text{s.t} \quad \mathbf{C}\widehat{\mathbf{Q}}=\mathbf{0}\, ,
\end{split}
\end{equation}
where $\widehat{\mathbf{Q}}_{o}$ is the original mode coefficient vector at the current time step; $\mathbf{C}=[\mathbf{C}_{R}, -\mathbf{C}_{I}] \in \mathbb{R}^{5\times2M}$.} 

 {Following the method of Lagrange multipliers, we obtain the conservative correction $\widehat{\mathbf{Q}_{c}}$
\begin{equation}\label{CorrectedQ}
    \widehat{\mathbf{Q}_{c}}=\left[\mathbf{I}-\mathbf{C}^{T}\left(\mathbf{C}\mathbf{C}^{T} \right)^{-1}\mathbf{C}\right]\widehat{\mathbf{Q}}_{o}\, ,
\end{equation}
where $\mathbf{I}$ is a $2M\times 2M$ identity matrix.}

 {Thus, in the temporal evolution, the above \emph{CONSERVE} (\ref{CorrectedQ}) and \emph{RECONSTRUCT} (\ref{Q_FS}) routines have to be implemented at every time step, e.g every intermediate step of the Runge-Kutta scheme that will be discussed in next section.}

%
%
%

\subsection{Time Discretization}

The high dimensionality and nonlinearity would make an implicit iterative time discretization really expensive. Thus, an explicit method is preferred.
Due to the diffusive nature of the collision operator, a stiff problem has to be solved,  and thus the corresponding stability condition forces the time step to be on the order of the square of the velocity step.
We will show this property in the following. The original proof is due to \cite{FPL_1x2v} and can easily extend to our spectral method.

What we need to solve is the following problem
\begin{equation}
\frac{d}{dt}\widehat{f}(\xi_{k})= F(\widehat{f}(\xi_{k}))\, ,
\end{equation}
where
\begin{equation}
 F(\widehat{f}(\xi_{k})) = \frac{1}{\varepsilon}\widehat{\textbf{Q}}(\widehat{f},\widehat{f})(\xi_{k})
\end{equation}
with $\widehat{\textbf{Q}}(\widehat{f},\widehat{f})$ defined in (\ref{Qhat_analytical}).

In practice, we employ a fourth-order explicit Runge-Kutta scheme that achieves high temporal accuracy and at the same time does not ruin the spectral accuracy.
Since the Runge-Kutta method is just a convex combination of first order Euler scheme, we only need to consider the first order Euler scheme
\begin{equation}\label{eqn:Euler_homogFPL}
\widehat{f}^{n+1}(\xi_{k}) = \widehat{f}^{n}(\xi_{k}) +\Delta t F(\widehat{f}^{n}(\xi_{k}))\, ,
\end{equation}
where the superscript $n$ denotes the mode value at the $n$-th time step.
The linear stability theory tells us the stability condition is determined by the eigenvalues of the Jacobian $\mathcal{J}_{k,l}=\frac{\partial F(\widehat{f}(\xi_{k}))}{\partial \widehat{f}(\xi_{l})}$. We need to
find an upper bound on the (negative) eigenvalues $\lambda$, such that $\lambda \Delta t < 1$.

Then, we have the following proposition
\newtheorem{prop0}{Proposition}
\begin{prop0}[Stability condition for homogeneous FPL]
For the first order Euler scheme, the time step $\Delta t$ should satisfy the following stability condition,
{\begin{equation}\label{eqn:Stability_homog_FPL}
\Delta t \leq 
 \frac{C\varepsilon L_{v}}{\|f\|_{L^{1}(\mathbb{R}^{3})}}\left( \frac{L_{v}}{N}\right)^{2}\,  ,
\end{equation} }
where $L_{v}$ is the lateral size of the fixed velocity domain, $\varepsilon$ is the Knudsen number and constant $C$ only depends on 
the computing domain $\Omega_{v}$. 
\end{prop0}

\newproof{pf}{Proof}
\begin{pf}
For the sake of generality, our proof works on general dimension $d$.
We rewrite (\ref{Qhat_analytical}) into two convolution forms
\begin{equation}\label{eqn:Qhat_2conv}
 \widehat{\textbf{Q}}(\widehat{f},\widehat{f}) (\xi) = \widehat{f} \ast G(\widehat{f}) (\xi) - \sum^{d}_{i,j=1} H_{i,j}(\widehat{f}) \ast J_{i,j}(\widehat{\textbf{S}};\widehat{f}) (\xi)
\end{equation}
with, $\xi=(\xi^{(1)}, \xi^{(2)}, \ldots, \xi^{(d)})$ being defined component-wisely,
\begin{equation}
 \begin{split}
  G(\widehat{f})(\xi) &:=  2\sqrt{\frac{2}{\pi}} \frac{R|\xi|-\sin(R|\xi|)}{R|\xi|} \widehat{f}(\xi) \, ; \\
 H_{i,j}(\widehat{f})(\xi) &:= \widehat{f}(\xi)\xi^{(i)}\xi^{(j)} \, ; \\
J_{i,j}(\widehat{\textbf{S}};\widehat{f})(\xi) &:= \widehat{f}(\xi)\widehat{\textbf{S}}_{i,j}(\xi) \, .
\end{split}
\end{equation}
The convolutions in (\ref{eqn:Qhat_2conv}) will be evaluated by the Trapezoidal quadrature rule, with the Fourier nodes $\widehat{f}(\xi_{k})$ being the quadrature points. That is,
\begin{equation}
 \widehat{\textbf{Q}}(\widehat{f},\widehat{f}) (\xi_{k}) = h^{d}_{\xi}\sum_{l}\omega_{l}\left[\widehat{f}(\xi_{k}-\xi_{l})G(\widehat{f})(\xi_{l})  - \sum^{d}_{i,j=1} H_{i,j}(\widehat{f})(\xi_{k}-\xi_{l})J_{i,j}(\widehat{\textbf{S}};\widehat{f})(\xi_{l}) \right]\, ,
\end{equation}
where $h_{\xi}$ is the step size in Fourier space as determined by (\ref{eqn:FFT_requirement}), and $\omega_{l}$ are quadrature weights.

According to \cite{FPL_1x2v}, the time step should satisfy
\begin{equation}
\Delta t \leq \frac{1}{\text{Lip}(F(\cdot))}\, ,
\end{equation}
where $\text{Lip}(F(\cdot))$ is the Lipschitz norm of $F(\cdot)$. This can be found through estimating the upper bound on the Jacobian
\begin{equation}
 \begin{split}\label{stab1}
|\mathcal{J}_{k,l}|&=\left|\frac{d}{d\widehat{f}(\xi_{l})}F(\widehat{f}(\xi_{k}))\right| \\
&\leq \frac{1}{\varepsilon}\frac{C}{L^{d}_{v}}\max\left(|\widehat{f}(\xi_{k}-\xi_{l})|, |\widehat{f}(\xi_{l})|\right) \\
&\ \ \ \cdot \left[\max_{\xi} \left| \frac{R|\xi|-\sin(R|\xi|)}{R|\xi|} \right| + |(\xi_{k}-\xi_{l})^{T}\widehat{\textbf{S}}(\xi_{l})(\xi_{k}-\xi_{l}) | + |\xi_{l}^{T}\widehat{\textbf{S}}(\xi_{k}-\xi_{l})\xi_{l} |\right] \\
&\leq \frac{C}{\varepsilon L_{v}}|\widehat{f}^{n}(0)|L^{2}_{\xi} \\
&\leq \frac{C}{\varepsilon L_{v}}\|f\|_{L^{1}(\mathbb{R}^{d})}\frac{1}{(\Delta v)^{2}}\, , \nonumber
\end{split}
\end{equation}
where the FFT relationship (\ref{eqn:FFT_requirement}) is applied, and it is not hard to observe the following uniform bound estimates
\begin{equation}
 |\widehat{\textbf{S}}(\xi)| \lesssim L^{d-1}_{v}\, , \quad |\xi^{T}\widehat{\textbf{S}}(\xi)\xi|\lesssim 1 \, , \quad |(\xi-w)^{T}\widehat{\textbf{S}}(\xi)(\xi-w)| \lesssim L^{d-1}_{v}L^{2}_{\xi} \, .
\end{equation}

{Therefore,  
the time step has to satisfy the stability condition
\begin{equation}
\Delta t \leq 
 \frac{C \varepsilon  L_{v}}{\|f\|_{L^{1}(\mathbb{R}^{d})} } \left(\frac{L_v}{N}\right)^2\, .
\end{equation}
for the constant $C$ depend only on the space dimension.
\\
Hence the stability condition  (\ref{eqn:Stability_homog_FPL}) holds.}

\end{pf}

\medskip

In practice, we employ a fourth-order explicit Runge-Kutta scheme and the conservation routine should be performed ad every intermediate step.
Recall our discretization of time $t_{n}=t_{0}+n\Delta t$, where $\Delta t$ is the time step size. Denote by $\mathbf{F}_{n}$ the distribution vector at time step $t_{n}$. In a time interval $[t_{n}, t_{n+1}]$, the numerical evolution $\mathbf{F}_{n}\rightarrow \mathbf{F}_{n+1}$ follows
\begin{equation}\label{RK4}
\begin{split}
&\widehat{\mathbf{F}_{n}}=\text{FFT}(\mathbf{F}_{n}), \qquad \ \widehat{\mathbf{K}^{1}_{n}}=\text{Compute}\left(\widehat{\mathbf{Q}}(\widehat{\mathbf{F}_{n}},\widehat{\mathbf{F}_{n}})\right), \, \widehat{\mathbf{K}^{1}_{n}}=\text{Conserve}(\widehat{\mathbf{K}^{1}_{n}}), \, \nonumber \\
& \qquad \mathbf{K}^{1}_{n}=\text{IFFT}\left( \widehat{\mathbf{K}^{1}_{n}} \right), \qquad\widetilde{\mathbf{F}}_{n}=\mathbf{F}_{n}+\Delta tK^{1}_{n}; \nonumber \\
& \widehat{\widetilde{\mathbf{F}}_{n}}=\text{FFT}(\widetilde{\mathbf{F}}_{n}), \qquad \   \widehat{\mathbf{K}^{2}_{n}}=\text{Compute}\left(\widehat{\mathbf{Q}}(\widehat{\widetilde{\mathbf{F}}_{n}},\widehat{\widetilde{\mathbf{F}}_{n}})\right), \, \widehat{\mathbf{K}^{2}_{n}}=\text{Conserve}(\widehat{\mathbf{K}^{2}_{n}}),  \nonumber \\
&\qquad\mathbf{K}^{2}_{n}=\text{IFFT}\left( \widehat{\mathbf{K}^{2}_{n}} \right),  \qquad\  \widetilde{\mathbf{F}}_{n}=\mathbf{F}_{n}+\frac{\Delta t}{2}K^{1}_{n}+\frac{\Delta t}{2}K^{2}_{n};  \\
&\widehat{\widetilde{\mathbf{F}}_{n}}=\text{FFT}(\widetilde{\mathbf{F}_{n}}), \qquad \  \, \widehat{\mathbf{K}^{3}_{n}}=\text{Compute}\left(\widehat{\mathbf{Q}}(\widehat{\widetilde{\mathbf{F}}_{n}},\widehat{\widetilde{\mathbf{F}}_{n}})\right), \, \widehat{\mathbf{K}^{3}_{n}}=\text{Conserve}(\widehat{\mathbf{K}^{3}_{n}}),  \nonumber \\
&\qquad\mathbf{K}^{3}_{n}=\text{IFFT}\left( \widehat{\mathbf{K}^{3}_{n}} \right), \qquad \  \widetilde{\mathbf{F}}_{n}=\mathbf{F}_{n}+\frac{\Delta t}{2}K^{1}_{n}+\frac{\Delta t}{2}K^{3}_{n}; \nonumber\\
&\widehat{\widetilde{\mathbf{F}}_{n}}=\text{FFT}(\widetilde{\mathbf{F}_{n}}), \qquad \   \widehat{\mathbf{K}^{4}_{n}}=\text{Compute}\left(\widehat{\mathbf{Q}}(\widehat{\widetilde{\mathbf{F}}_{n}},\widehat{\widetilde{\mathbf{F}}_{n}})\right), \, \widehat{\mathbf{K}^{4}_{n}}=\text{Conserve}(\widehat{\mathbf{K}^{4}_{n}}), \nonumber \\
&\qquad  \mathbf{K}^{4}_{n}=\text{IFFT}\left( \widehat{\mathbf{K}^{4}_{n}} \right),  \qquad\ \ \mathbf{F}_{n+1}=\mathbf{F}_{n}+\frac{1}{6}( 3\mathbf{K}^{1}_{n}+\mathbf{K}^{2}_{n}+\mathbf{K}^{3}_{n}+\mathbf{K}^{4}_{n} ).
\end{split}
\end{equation}
where $\widetilde{\mathbf{F}}_{n}$ a generic intermediate step; IFFT is the (discrete) fast inverse Fourier transform routine.

\section{The RKDG Method for Vlasov-Poisson Equation}\label{sec:RKDG_VP}
The VP system is a nonlinear kinetic system modeling the transport of charged particles
in a collisionless plasma, under the effect of a self-consistent electrostatic field and possibly
an externally supplied field. The electrostatic potential is coupled through Poisson equation. The collisionless VP exhibits a variety of dynamical phenomena, for example, the well-known filamentation (filaments in phase space and steep gradients in $v$) and Landau damping.

With coupling to Poisson equation, the collisionless Vlasov Poisson problem becomes
\begin{align}\label{VP_eqn}
&\ \text{The VP (Collisionless) Problem} \nonumber \\
&\  \nonumber \\
&\quad \ \partial_{t} g(x, v, t) + v \cdot \bigtriangledown_{x} g(x, v, t) - \mathbf{E}(t,x)\cdot\nabla_{v}g  =  0 \, ,   
 \nonumber \\
&\quad \ \mathbf{E}(t,x)=-\nabla_{x} \Phi(t,x)\, , \nonumber \\
&\quad \ \Delta_{x}\Phi(t,x)=\int_{R^{3}}g(t,x,v) dv -1  \, ,       \qquad \ \text{for}\ \  (x,v)\in \Omega_x \times \mathbb{R}_v\\
&\quad \ \Phi(t,x)=\Phi_{B}(t,x) \quad x\in \partial\Omega_{x} \, , \nonumber \\
&\quad \ g(0, x, v) = f_{n}(x, v) \, ,\nonumber
\end{align}
where $f_{n}$ is the current solution of the homogeneous Landau equation.

\subsection{The Semi-discrete DG Form}\label{sec:semiDG}

In this section, we introduce a conservative Runge-Kutta Discontinuous Galerkin (RKDG) scheme for the VP equation (\ref{VP_eqn}), for $(x,v)\in \Omega=\Omega_{x}\times\Omega_{v}\subseteq \mathbb{R}^{+}\times \mathbb{R}^{d}$. Or, we restrict the problem to the first spatial dimension $\mathbf{x}=(x,0,0)$, $\mathbf{E}=(E,0,0)$. The conservation properties are proved to be well satisfied if we choose a piecewise polynomial approximation space covering $d+2$ collision invariants.

We first list some notations for the DG method in use. Consider the computing domain $\Omega=\Omega_{x}\times\Omega^E_{v}=[0,L_{x}]\times[- L^E_{v},L^E_{v}]^{3}$, 1D in $x$-space and 3D in $v$-space. 

In this case, the cut-off domain  $\Omega^E_{v}$ in velocity space now depends on the electric field $E(x)=-\nabla \Phi$ according to the mean-field Vlasov-Poisson flow in \eqref{VP_eqn}. The particular choice of diameter constant $L^E_v$ is chosen by taking
 \begin{align}\label{L^E}
L^E_v = L_0 +  cE^* ; \qquad \qquad \text {with}\ \       E^*= \max_{x\in\Omega_{x}} |E_n(x)|
\end{align}
where  $E_n(x) \!= \!\int_0^x \! \int_{\mathbb{R}_v} \!f_n(x,v)dvdx \!-\!1$, and the factor $c$ is of order of unity.  Thus $E^*  \!\le \!\int_{\Omega_{x}\times\mathbb{R}_v} \!\! f_n(x,v) dv\, dx \! + \!1.$
Heuristically, this cut-off domain correction in $v$ space allows for the computational solution $g(x,v,t)$  of the Vlasov flow along the Hamiltonian characteristic fields  at each time step, given by $(x-vt^n, v-E_{n-1}(x)\, t^n)$ to keep its initial support transported by the characteristic curves, well inside the computational domain $\Omega^{E^*}_{v}$ with $L^E_v = L_0 +  cE^*$.
We stress that this approach works for periodic boundary conditions in $x$-space set on $\Omega_x$. It results in a uniform in time $L^E_v$, since the solution associated to the Vlasov Poisson system in one dimension in $x$-space yields global uniformly bounded 
electric fields. That means the set $\Omega^E_{v}$ does not need to be updated with the time step evolution.

Denote by $\mathcal{T}^{x}_{h}$ and
$\mathcal{T}^{v}_{h}$ the regular partitions of $\Omega_{x}$ and $\Omega_{v}$, respectively, with
\begin{eqnarray*}
\mathcal{T}^{x}_{h}\!\! &=&\!\!\bigcup^{N_{x}}_{1} I_{i}=\bigcup^{N_{x}}_{1}[x_{i-1/2},x_{i+1/2}) \\
\mathcal{T}^{v}_{h}\!\! &=&\!\!\bigcup^{N^{3}_{v}}_{|j|=1}K_{j}\!\!=\!\!\bigcup^{N_{v}}_{j_1,j_2,j_3=1} [v_{j_1-1/2},v_{j_1+1/2})\times[v_{j_2-1/2},v_{j_2+1/2})\times[v_{j_3-1/2},v_{j_3 +1/2})\, ,
\end{eqnarray*}
with $x_{1/2}=0$, $x_{N_{x}+1/2}=L_{x}$, $v_{1/2}=-L_{v}$ and $v_{N_{v}+1/2}=L_{v}$.

Then, $\mathcal{T}_{h}=\{E: E=I_{x}\times K_{v}, \forall I_{x}\in \mathcal{T}^{x}_{h},\forall K_{v}\in\mathcal{T}^{v}_{h}\}$ defines a partition of $\Omega$.
Denote by $\varepsilon_{x}$ and $\varepsilon_{v}$ be set of edges of $\mathcal{T}^{x}_{h}$ and $\mathcal{T}^{v}_{h}$, respectively. Then, the edges of $\mathcal{T}_{h}$
will be $\varepsilon = \{I_{x}\times e_{v}: \forall I_{x}\in \mathcal{T}^{x}_{h}, \forall e_{v}\in \varepsilon_{v}\}\cup \{e_{x}\times K_{v}: \forall e_{x} \in \varepsilon_{x}, \forall K_{v}\in \mathcal{T}^{v}_{h} \}$. In addition, $\varepsilon_{x}=\varepsilon^{i}_{x}\cup \varepsilon^{b}_{x}$ with $\varepsilon^{i}_{x}$ and $\varepsilon^{b}_{x}$ being the interior and boundary edges, respectively. Same for velocity domain.
The mesh size $h=\max(h_{x},h_{v})=\max_{E\in \mathcal{T}_{h}}\text{diam}(E)$, with $h_{x}=\max_{I_{x}\in \mathcal{T}^{x}_{h}}\text{diam}(I_{x})$ and
$h_{v}=\max_{K_{v}\in \mathcal{T}^{v}_{h}}\text{diam}(K_{v})$.

Next, we define the following approximation space (note that we only have 1D in $x$):
\begin{eqnarray}\label{eqn:RKDG_Xspace}
X^{l}_{h}=\{f\in L^{2}_{\Omega}: g|_{E} \in P^{l}(I_{x})\times P^{l}(K_{v}), \forall E=I_{x}\times K_{v}\in \mathcal{T}_{h}\}\, ,
\end{eqnarray}
and
\begin{eqnarray}\label{eqn:RKDG_Wspace}
W^{l}_{h}=\{f\in L^{2}_{\Omega}: g|_{E} \in P^{l}(I_{x})\times Q^{l}(K_{v}), \forall E=I_{x}\times K_{v}\in \mathcal{T}_{h}\}\, ,
\end{eqnarray}
where $P^{l}(K)$ denotes the space of polynomials of total degree at most $l$ on some element $K$, while $Q^{l}$ the space of polynomials of degree $l$ in each variable on $K$.
$P^{l}(K)$ has number of degrees of freedom $(l+1)^{d}$, while $Q^{l}(K)$ has degrees of freedom $\tiny \sum^{l}_{i=0}\binom{i+d-1}{d-1}$ (here $d=3$).

Since basis polynomials are piecewise defined over each element, we need to introduce the concepts of jumps and averages. For any test function $\phi_{h}(x,v)\in X^{l}_{h}$ (or, $W^{l}_{h}$),
define $(\phi_{h})^{\pm}_{i+1/2,v}=\lim_{\epsilon\rightarrow 0}\phi_{h}(x_{i+1/2}\pm \epsilon, v)$, $(\phi_{h})^{\pm}_{x, K_{v}}=\phi_{h}|_{K^{\pm}_{v}}$.
For any edge $e_{x}\in \varepsilon_{x}$, which is actually one end point of intervals, and any edge $e_{v}\in \varepsilon_{v}$, with $\mathbf{n}_{v}^{\pm}$ as the outward unit normal to $\partial K^{\pm}_{v}$,
the jumps across $e_{x}$ and $e_{v}$ are defined as
\begin{equation}
[\phi_{h}]_{x_{i}}=(\phi_{h})^{+}_{i-1/2,v}-(\phi_{h})^{-}_{i-1/2,v}, \quad [\phi_{h}]_{v}=(\phi_{h})^{+}_{x, K_{v}}\mathbf{n}^{+}_{v} + (\phi_{h})^{-}_{x, K_{v}}\mathbf{n}^{-}_{v}\, .
\end{equation}

and the averages are
\begin{equation}
\{\phi_{h}\}_{x_{i}}=\frac{1}{2}((\phi_{h})^{+}_{i-1/2,v}+(\phi_{h})^{-}_{i-1/2,v}), \quad \{\phi_{h}\}_{v}=\frac{1}{2}((\phi_{h})^{+}_{x,K_{v}}+(\phi_{h})^{-}_{x,K_{v}})\, .
\end{equation}

Here and below, we denote by $E_{h}$ the discrete electric field computed from the Poisson equation. With proper partitioning,
we can assume each direction of $v$ has a single sign.

The DG scheme for the nonlinear VP equation is described as follows. We seek an approximate solution $g_{h}(x,v)\in X^{l}_{h}$ (or $W^{l}_{h}$)
such that, for any test function $\phi_{h}(x,v)\in X^{l}_{h}$ (or $W^{l}_{h}$)
\begin{equation}\label{SemiDG_VP}
\int_{I_{i}\times K_{j}} (g_{h})_{t}\varphi_{h}dxdv = H_{i,j}(g_h, E_h, \varphi_h)
\end{equation}
where
\begin{align}
&H_{i,j}(g_h, E_h,\varphi_h)  \nonumber \\
&\quad=\int_{I_{i}\times K_{j}}v_1 g_{h}(\varphi_{h})_{x}dxdv - \int_{K_j}(\widehat{v_1 g_{h}}\varphi^{-}_{h})_{i+\frac{1}{2},v}dv + \int_{K_j}(\widehat{v_1 g_{h}}\varphi^{+}_{h})_{i-\frac{1}{2},v}dv  \nonumber \\
&\quad\ \ -\int_{I_{i}\times K_{j}}E_{h}g_{h}\partial_{v_1}\varphi_{h}dxdv + \int_{I_i}\int_{\varepsilon_{v}}(\widehat{E_{h}g_{h}}\varphi^{-}_{h})_{x,j_1+\frac{1}{2}}ds_{v}dx \\
&\qquad   -\ \int_{I_i}\int_{\varepsilon_{v}}(\widehat{E_{h}g_{h}}\varphi^{+}_{h})_{x,j_1-\frac{1}{2}}ds_{v}dx \, . \nonumber
\end{align}
Here, $j=(j_1, j_2, j_3)$ is the multi-index, corresponding to the three directions of $v$. The following upwinding fluxes (the trace at the element interfaces) are used,
\begin{equation}
\widehat{v_1 g_h}=\left\{ \begin{array}{ll}
                                 v_1 g^{-}_h, & where\hbox{if $v_1 \geq 0$ in $K_j$,} \\
                                 v_1 g^{+}_h, & \hbox{if $v_1 < 0$ in $K_j$.}
                               \end{array}
                             \right.
\end{equation}

and
\begin{equation}
\widehat{E_h g_h}=\left\{ \begin{array}{ll}
                                 E_h g^{-}_h, & \hbox{if $\int_{I_i}E_{h}dx \leq 0$,} \\
                                 E_h g^{+}_h, & \hbox{if $\int_{I_i}E_{h}dx > 0$.}
                               \end{array}
                             \right.
\end{equation}

The electric field is solved from the Poisson's equation, as is used in \cite{CGM_RKDG_VP}. In the one-dimensional case, the exact solution of the Poisson's equation can be obtained
through the classical representation of Green's function,
if we enforce the periodicity condition $\Phi(0)=\Phi(L_{x})$,
\begin{equation}\label{eqn:ExactPoisson_phi}
\Phi_h=\int^{x}_{0}\int^{s}_{0}\rho_{h}(z,t)dzdx-\frac{x^{2}}{2}-C_{E}x \, ,
\end{equation}
where $\rho_{h}=\int_{\Omega_v}g_{h}dv$, $C_{E}=-\frac{L_{x}}{2} + \frac{1}{L_{x}}\int^{L_x}_{0}\int^{s}_{0}\rho_{h}(z,t)dzds$, and
\begin{equation}\label{ExactE_VP}
E_{h}=-\Phi'=C_{E}+x-\int^{x}_{0}\rho_{h}(z,t)dz \, .
\end{equation}

The above semi-DG problem (\ref{SemiDG_VP}) can be solved by coupling with a suitable time discretization, e.g. total variation diminishing (TVD) Runge-Kutta method.
The third order TVD-RK method for evolving $t_{n}\rightarrow t_{n+1}$ is implemented as
\begin{eqnarray}\label{RK3_VP}
\int_{I_i\times K_j}g^{(1)}_{h}\varphi_h dxdv &=& \int_{I_i\times K_j}g^{n}_{h}\varphi_h dxdv + \Delta t H_{i,j}(g^{n}_h, E^{n}_h, \varphi_h)\, , \nonumber \\
\int_{I_i\times K_j}g^{(2)}_{h}\varphi_h dxdv &=& \frac{3}{4}\int_{I_i\times K_j}g^{n}_{h}\varphi_h dxdv + \frac{1}{4}\int_{I_i\times K_j}g^{(1)}_{h}\varphi_h dxdv \nonumber\\
&\ &\ \ + \frac{\Delta t}{4} H_{i,j}(g^{(1)}_h, E^{(1)}_h, \varphi_h)\, ,  \\
\int_{I_i\times K_j}g^{n+1}_{h}\varphi_h dxdv &=& \frac{1}{3}\int_{I_i\times K_j}g^{n}_{h}\varphi_h dxdv + \frac{2}{3}\int_{I_i\times K_j}g^{(2)}_{h}\varphi_h dxdv \nonumber\\
&\ &\ \ + \frac{2\Delta t}{3} H_{i,j}(g^{(2)}_h, E^{(2)}_h, \varphi_h)\, ,\nonumber \\
\end{eqnarray}
where $E^{(1)}_h,E^{(2)}_h$ are also obtained through the exact representation (\ref{ExactE_VP}). Readers can refer to \cite{Shu_TVDRK} for a detailed introduction to TVD Runge-Kutta methods.

This completes the RKDG scheme for nonlinear VP problem. We propose to apply basis function $\varphi_{h}|_{K_{j}}=1,v,|v|^{2}$, as is proposed in study of Vlasov-Maxwell equations in \cite{CGLM_DGVM}, hoping that the RKDG scheme can well preserve mass, momentum and energy.

\noindent\textbf{Remark. }
{\it To ensure a positive DG solution, many authors have successfully applied positivity-preserving limiters in the intermediate time steps. Please refer to \cite{ZhangShu_positivitylimiter0,ZhangShu_positivitylimiter1,ZhangShu_positivitylimiter2,ZhangShu_positivitylimiter3,CGP_Positive} for full
descriptions and applications. We summarize the scheme here. For each intermediate step of Runge-Kutta method,

\begin{itemize}
 \item On each mesh element $E_{i,j}=I_{i}\times K_{j}$, compute $T_{i,j}:=\min_{(x,v)\in S_{i,j}} g_{h}(x,v)$, where $S_{i,j}=\big(S^{x}_{i}\otimes \hat{S}^{v}_{j} \big) \cup \big(\hat{S}^{x}_{i}\otimes S^{v}_{j} \big)$,
and $S^{x}_{i},S^{v}_{j}$ are sets of $(l+1)$ Gauss quadrature points and $\hat{S}^{x}_{i},\hat{S}^{v}_{j}$ sets of $(l+1)$ Gauss-Lobatto quadrature points.
\item Compute $\widetilde{f}_{h}(x,v)=\theta\big( g_{h}(x,v)-(\overline{g_{h}})_{i,j}\big)+ (\overline{g_{h}})_{i,j}$ with $(\overline{g_{h}})_{i,j}$ the average over element $E_{i,j}$ and $\theta=\min\{1, |(\overline{g_{h}})_{i,j}|/|T_{i,j}-(\overline{g_{h}})_{i,j}|\}$.
\item Update $g_{h}\leftarrow\widetilde{g_{h}}$.
\end{itemize}

The above limiter adjusts the function to be positive while preserving the cell average. Thus, application of such positivity-preserving limiter still achieves conservation of total mass, yet however will deteriorate the conservation of energy. This limiter may be added when necessary, but for the time being, we would like to highlight the conservation of all desired moments.
}

\subsection{Conservation and $L^{2}$-Stability}\label{sec:DG_L2_prof}

A piecewise polynomial approximation subspace containing all collision invariants will be applied. We will show the total mass (charge) and momentum is conserved, up to some boundary error terms; as for the total energy, the variation relies on
the approximation accuracy of the solution together with the projection error of the potential $\Phi_h$. Also, the approximate solution is $L^{2}$ stable. The following propositions are extensions of some related results studied in \cite{Heath_dissertation, CGM_RKDG_VP, CGLM_DGVM} in higher dimensions.

\newtheorem*{prop1}{Proposition}
\begin{prop1}[Conservations of total mass and momentum]
The approximate solution $g_{h} \in X^{l}_{h}$ (or, $W^{l}_{h}$) for semi-DG problem (\ref{SemiDG_VP}) satisfies
\begin{equation}\label{eqn:VP_masscons}
\frac{d}{dt}\int_{\mathcal{T}_{h}}g_{h}dxdv = \Theta_{h,1}(g_{h},E_{h})\, ,
\end{equation}
with
\begin{equation}
\Theta_{h,1}(g_{h},E_{h}) = \int_{\mathcal{T}^{x}_{h}}\int_{\varepsilon^{b}_{v}}(\widehat{E_{h}g_{h}})_{x,N_{v}+\frac{1}{2}}ds_{v}dx - \int_{\mathcal{T}^{x}_{h}}\int_{\varepsilon^{b}_{v}}(\widehat{E_{h}g_{h}})_{x,\frac{1}{2}}ds_{v}dx \, ,
\end{equation}

and
\begin{equation}\label{eqn:VP_momcons}
\frac{d}{dt}\int_{\mathcal{T}_{h}}g_{h}vdxdv = \Theta_{h,2}(g_{h},E_{h})\, ,
\end{equation}
with
\begin{equation}
\Theta_{h,2}(g_{h},E_{h}) = \int_{\mathcal{T}^{x}_{h}}\int_{\varepsilon^{b}_{v}}(\widehat{E_{h}g_{h}}v)_{x,N_{v}+\frac{1}{2}}ds_{v}dx - \int_{\mathcal{T}^{x}_{h}}\int_{\varepsilon^{b}_{v}}(\widehat{E_{h}g_{h}}v)_{x,\frac{1}{2}}ds_{v}dx\, .
\end{equation}
Here, boundary error terms $\Theta_{h,1}(g_{h},E_{h})$ and $\Theta_{h,2}(g_{h},E_{h})$ are negligible if $\Omega_{v}$ is selected according to the criteria \eqref{L^E} discussed in the previous subsection. 
\end{prop1}

\newproof{pf1}{Proof}
\begin{pf1}
Take $\varphi_h=1$, then
\begin{equation}
\begin{split}
\sum_{i,j}H_{i,j}(g_h, E_h,1) &=  \int_{\mathcal{T}^{v}_{h}}\int_{\varepsilon_{x}}\widehat{v_1 g_{h}}[1]_{x}ds_{x}dv - \int_{\mathcal{T}^{x}_{h}}\int_{\varepsilon_{v}}\widehat{E_{h}g_{h}}[1]_{v_{1}}ds_{v}dx \\
& = \int_{\mathcal{T}^{x}_{h}}\int_{\varepsilon^{b}_{v}}(\widehat{E_{h}g_{h}})_{x,N_{v}+\frac{1}{2}}ds_{v}dx - \int_{\mathcal{T}^{x}_{h}}\int_{\varepsilon^{b}_{v}}(\widehat{E_{h}g_{h}})_{x,\frac{1}{2}}ds_{v}dx\, .
\end{split}
\end{equation}
where the periodicity in $x$ is considered.

Take $\varphi_h=v_{1}$, then
\begin{align*}
\sum_{i,j}H_{i,j}(g_h, E_h,v_{1}) &=  \int_{\mathcal{T}^{v}_{h}}\int_{\varepsilon_{x}}\widehat{v_1 g_{h}}[v_{1}]_{x}ds_{x}dv \ -\ \int_{\mathcal{T}^{x}_{h}}\int_{\mathcal{T}^{v}_{h}}E_{h}g_{h}dxdv \\
 &\  - \ \int_{\mathcal{T}^{x}_{h}}\int_{\varepsilon_{v}}\widehat{E_{h}g_{h}}[v_{1}]_{v_{1}}ds_{v}dx\, .
\end{align*}
The first term above is zero due to the periodic boundary conditions; the third term is the boundary error same as above; let's only look at the
second term. Thanks to the exact solver for Poisson equation (\ref{eqn:ExactPoisson_phi}) and (\ref{ExactE_VP}),
\begin{align*}
\int_{\mathcal{T}^{x}_{h}}\int_{\mathcal{T}^{v}_{h}}E_{h}g_{h}dxdv = \int_{\mathcal{T}^{x}_{h}}\rho_{h}E_{h}dx = -\int_{\mathcal{T}^{x}_{h}}E_{h}(E_{h})_{x}dx + \int_{\mathcal{T}^{x}_{h}}E_{h}dx = 0\, .
\end{align*}

The cases for $\varphi_h=v_{2}$ and $\varphi_h=v_{3}$ follow the same way.
\end{pf1}

\newtheorem*{prop2}{Proposition}
\begin{prop2}[Variation of total energy]
The total energy of the approximate solution $g_{h} \in X^{l}_{h}$ (or $W^{l}_{h}$) for the semi-DG problem (\ref{SemiDG_VP}) satisfies
\begin{align}\label{eqn:VP_totene}
\frac{d}{dt}\left(\frac{1}{2}\int_{\mathcal{T}_{h}}g_{h}|v|^{2}dxdv +  \frac{1}{2}\int_{\mathcal{T}^{x}_{h}}|E_{h}|^{2}dx\right)
 &= \Theta_{h,3}(g_{h},E_{h})\\
&= \Theta_{h,3}(g_{h}-g,\Phi_{h}-\mathbf{P}\Phi_{h})\, ,\nonumber
\end{align}
with
\begin{equation*}
\Theta_{h,3}(g_{h},E_{h}) = \int_{\mathcal{T}_{h}}(\Phi_{h})_{x}g_{h}v_{1}dxdv - \int_{\mathcal{T}_{h}}\Phi_{h}(g_{h})_{t}dxdv\, ,
\end{equation*}
where $\mathbf{P}\Phi_{h}$ is the projection of $\Phi_{h}$ onto $X^{l}_{h}$ (or, $W^{l}_{h}$) and $\mathbf{P}\Phi_{h} = \Phi_{h}$ on all interfaces of $\mathcal{T}^{x}_{h}$ (such that $\mathbf{P}\Phi_{h}$ is continuous).
\end{prop2}

\newproof{pf2}{Proof}
\begin{pf2}
Take $\varphi_h=\frac{1}{2}|v|^{2}$, then
\begin{equation*}
\begin{split}
\sum_{i,j}H_{i,j}(g_h, E_h,\frac{1}{2}|v|^{2}) &=  \int_{\mathcal{T}^{v}_{h}}\int_{\varepsilon_{x}}\widehat{v_1 g_{h}}\frac{1}{2}[|v|^{2}]_{x}ds_{x}dv - \int_{\mathcal{T}^{x}_{h}}\int_{\varepsilon_{v}}\widehat{E_{h}g_{h}}\frac{1}{2}[|v|^{2}]_{v_{1}}ds_{v}dx  \\
&-\int_{\mathcal{T}_{h}}E_{h}g_{h}\partial_{v_1}\varphi_{h}dxdv\, .
\end{split}
\end{equation*}
The first term above is zero due to the periodicity; the second term is the boundary error, which is zero if we assume the solution is compactly supported in $\Omega_{v}$.

On the other hand, noticing again the exact Poisson solver (\ref{eqn:ExactPoisson_phi}) and (\ref{ExactE_VP}),
\begin{equation}
\frac{1}{2}\int_{\mathcal{T}^{x}_{h}}|E_{h}|^{2}dx = -\int_{\mathcal{T}_{h}}\Phi_{h}(g_{h})_{t}dxdv\, ,
\end{equation}
which gives (\ref{eqn:VP_totene}).

If we take $\varphi_h = \mathbf{P}\Phi_{h} \in X^{l}_{h}$ (or $W^{l}_{h}$), then we obtain $\Theta_{h,3}(g_{h},\mathbf{P}\Phi_{h})=0$, which is also
valid for the exact solution $g$. The exact solution $g$ also obviously conserves total energy, which implies $\Theta_{h,3}(g,\Phi_{h}-\mathbf{P}\Phi_{h})=0$. Thus, $\Theta_{h,3}(g_{h},E_{h})= \Theta_{h,3}(g_{h}-g,\Phi_{h}-\mathbf{P}\Phi_{h})$.
\end{pf2}

This proposition means the variation of total energy relies on the numerical error of $g-g_{h}$ and projection error $\Phi_{h}-\mathbf{P}\Phi_{h}$.
If the Poisson equation is not solved by exact formula but instead through a local DG method, then with special choice of flux, the total energy on the discrete level is proven to be conserved, see \cite{CGLM_DGVM} in the case of periodic boundary condition with a large enough domain depending on the initial data. But here, we focus on the inhomogeneous model coupled with the Landau collision operator, thus the exact Poisson solver is preferred without many extra efforts. Actually, when a relatively fine DG mesh is applied, the variations on total energy are negligible.

\newtheorem*{prop3}{Proposition}
\begin{prop3}[$L^{2}$-stability]
The approximate solution $g_{h} \in X^{l}_{h}$ (or $W^{l}_{h}$) for semi-DG problem (\ref{SemiDG_VP}) decays enstrophy
\begin{equation}
\frac{d}{dt}\int_{\mathcal{T}_{h}}g^{2}_{h}dxdv = \Theta_{h,4}(g_{h},E_{h})\leq 0\, ,
\end{equation}
with
\begin{equation}
\Theta_{h,4}(g_{h},E_{h}) = -\frac{1}{2}\int_{\mathcal{T}^{v}_{h}}\int_{\varepsilon_{x}} |v_{1}|[g_{h}]^{2}_{x}ds_{x}dv -\frac{1}{2}\int_{\mathcal{T}^{x}_{h}}\int_{\varepsilon_{v}} |E_{h}|[g_{h}]^{2}_{v_{1}}ds_{v}dx\, .
\end{equation}
\end{prop3}

\newproof{pf3}{Proof}
\begin{pf3}
Take $\varphi_h=g_{h}$, then
\begin{equation}
\begin{split}
\sum_{i,j}H_{i,j}(g_h, E_h,g_{h}) &=  \int_{\mathcal{T}_{h}}v_{1}g_{h}(g_{h})_{x}dxdv + \int_{\mathcal{T}^{v}_{h}}\int_{\varepsilon_{x}}\widehat{v_1 g_{h}}[g_{h}]_{x}ds_{x}dv \\
&-\int_{\mathcal{T}_{h}}E_{h}g_{h}(g_{h})_{v_{1}}dxdv
- \int_{\mathcal{T}^{x}_{h}}\int_{\varepsilon_{v}}\widehat{E_{h}g_{h}}[g_{h}]_{v_{1}}ds_{v}dx  \\
&:= a_{1} + a_{2}\, ,
\end{split}
\end{equation}
where, noticing the definition of upwinding flux
\begin{equation}
\widehat{v_1 g_{h}} = \{v_1 g_{h}\}_{x} - \frac{|v_{1}|}{2}[g_{h}]_{x}\, ,
\end{equation}
one can easily obtain,
\begin{equation}
\begin{split}
a_{1} &=  \int_{\mathcal{T}_{h}}v_{1}g_{h}(g_{h})_{x}dxdv + \int_{\mathcal{T}^{v}_{h}}\int_{\varepsilon_{x}}\widehat{v_1 g_{h}}[g_{h}]_{x}ds_{x}dv \\
&= -\frac{1}{2}\int_{\mathcal{T}^{v}_{h}}\int_{\varepsilon_{x}}|v_{1}|[g_{h}]^{2}_{x}ds_{x}dv\, ,
\end{split}
\end{equation}
and similarly
\begin{equation}
\begin{split}
a_{2} &=  -\int_{\mathcal{T}_{h}}E_{h}g_{h}(g_{h})_{v_{1}}dxdv
- \int_{\mathcal{T}^{x}_{h}}\int_{\varepsilon_{v}}\widehat{E_{h}g_{h}}[g_{h}]_{v_{1}}ds_{v}dx \\
&=-\frac{1}{2}\int_{\mathcal{T}^{x}_{h}}\int_{\varepsilon_{v}}|E_{h}|[g_{h}]^{2}_{v_{1}}ds_{v}dx\, .
\end{split}
\end{equation}
So, $\Theta_{h,4}(g_{h},E_{h})=a_{1}+a_{2}\leq 0$.

\end{pf3}

\section{The Linking Process - Conservative Projection}\label{sec:linking}
So far, we have solved two subproblems separately: Vlasov-Poisson equation and homogeneous Landau equation. The next step is to link
them together, i.e project the Fourier series solution of the homogeneous Landau equation onto the DG mesh. If denote by $F_{n}(f)$ the Fourier series solution of the homogeneous Landau equation at the $n$-th time step, and $P:L^{2}(\Omega_{v})\rightarrow X^{l}_{h}$ (or, $W^{l}_{h}$) the $L^{2}$ projection, then according to the time splitting scheme, the initial condition for $(n+1)$-st Vlasov-Poisson problem (\ref{eqn:timesplitting_vp}) is
\begin{equation}
g(0,x,v)=P(F_{n}(f))(x,v)\, .
\end{equation}

 {Under a pure homogeneous setting, the conservation routine specially designed in \cite{GT_jcp} should suffice to complete a conservative spectral solver for homogeneous Landau equation. However, in the time-splitting framework, after being projected to DG space, conservation of desired moments will be broken if the conservation routine still correct the collision operator on discrete level of Fourier modes. Thus, such a concern inspired us to develop the novel conservation routine in Section \ref{Sec:ConserveRoutine_FPL}. This conservation routine corrects moments in Fourier space rather than in velocity space. It can be easily claimed that the conservation correction ensures conservation of the whole inhomogeneous FPL system being treated through a time-splitting scheme.}

 {Indeed, the crucial idea is to conserve all moments of interest (here, mass, momentum and kinetic energy) independent of quadrature rules, that is, conserves moments fully on level of piecewise polynomials in the DG space $X^{l}_{h}$ (or, $W^{l}_{h}$). This is achieved automatically even one projects the Fourier mode solution to DG space, since our conservation correction is reasoned from \eqref{eqn:generalConserve} and our polynomial basis space is exactly spanned by all collision invariants, i.e $1, v, |v|^{2}$. }

 {In DG space, the $L^{2}$ conservation properties proven in section \ref{sec:DG_L2_prof} continue to guarantee
a conservative solution. When going back from DG space to Fourier
space, the Fourier modes are always corrected by \eqref{CorrectedQ} which automatically satisfies \eqref{eqn:generalConserve}, which is independent of the selection of Fourier modes.
Therefore, in a word, in an inhomogeneous setting and under time-splitting framework, the new conservation routine in Section \ref{Sec:ConserveRoutine_FPL} will replace the one
specially designed for homogeneous case \cite{GT_jcp}, and ensure the conservation property always hold.}

\noindent\textbf{Remark. }
\textit{We expect the whole discrete scheme to be stable and also to be able to construct a priori error estimates. These two goals will be done in a future project.
}

\section{Parallelization}
One common feature for nearly all realistic kinetic models is the high dimensionality. Plus the higher than linear complexity, it addresses the importance of implementations of parallel computing.

For RKDG schemes for VP problem, the parallelization becomes more natural due to the locality of basis functions. Once all the nodes can access to the information from previous time step, the evolution of each grid point
is done independently without communications across computing nodes. After evolution is done for the current time step for all nodes, the information will be gathered together and redistributed to all computing nodes in the
community. We will use Message Passing Interface (MPI) \cite{MPI} to distribute the velocity grid points.

Next, at each time step $t_k$ and a fixed space grid point $x_i$,  
the spectral solver used for solving the homogeneous FPL equation only ``sees'' the particles at position $(x_i,v_j, t_k)$ for any $0<j <N_v$ the same spatial grid point through the collision term. 
Since collisions involve all participating velocity grid points then, in
order to avoid large amount of communicating latency, we only distribute spatial grid points across the 
computing node community, and thus restrict all of the needed information at the current $k$-time step on the same computing node.


To further parallelize the computation, we realize that, for each phase velocity grid point $\xi$, the computation of $\widehat{Q}(\xi)$ is a weighted sum over all phase velocities $w$, with no
information interrupted by other grid points $\xi$'s. Similar features also apply to the integrations in RKDG method for the VP problem. Thus, the work load will be further shared using OpenMP \cite{Openmp}.

As the majority of computations occur in the collision steps, the computational time consumed in collisions will dominate. Since all information needed for collisions will be kept on the same computing node and only spatial grid points are distributed, an almost linear strong scaling efficiency would be expected. We run tests on a typical linear Landau damping problem for the Landau-Poisson system, and record the time consumed for one single time step in Table \ref{table:para_inhomoFPL}. This example is associated with the one in Figure \ref{fig:damping_k05}.
Tests run on Xeon Intel 3.33GHz Westmere processors (on cluster Lonestar-TACC \cite{TACC}).
\begin{table}[!ht]
\centering
\begin{tabular}{|c c c|}
  \hline
   nodes & cores & wall clock time (s)  \\
   \hline
  1 & 12 & 1228.18 \\
  2& 24 & 637.522  \\
  4& 48 &  307.125 \\
  8 & 96 & 154.385  \\
  16& 192 & 80.6144 \\
  32& 384 & 41.314  \\
\hline
\end{tabular}
\caption{The wall clock time for one single time step of a typical linear Landau damping problem. }\label{table:para_inhomoFPL}
\end{table}

\subsection{Numerical Results and Applications}\label{sec:results}

\subsubsection{Single Species Charge Carriers}\label{sec:numApp_singleSpecies}

This example is to validate our conservative solver for in the homogeneous setting. We test our scheme to a sum of two Gaussians in 3D velocity space, to compute the evolution of entropy and moments and thus verify its validity,
\begin{equation}
f_{0}(v)=\frac{1}{2(2\pi\sigma^{2})^{3/2}}\left[ \text{exp}\left(-\frac{|v-2\sigma e|^{2}}{2\sigma^{2}} \right) + \text{exp}\left(-\frac{|v+2\sigma e|^{2}}{2\sigma^{2}} \right) \right]\, ,
\end{equation}
with parameter $\sigma=\pi/10$ and $e=(1,0,0)$.

We select domain $\Omega_{v}=[-3,3]^{3}$, number of modes in each direction $N=32$.
\begin{figure}[!htb]
\begin{minipage}[t]{0.45\linewidth}
\centering
\includegraphics[width=70mm]{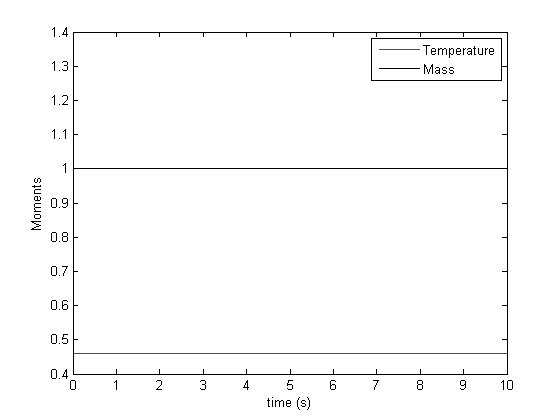}
\caption{The evolution of moments of numerical solution}\label{Evo_Pdf_FPL_2Gauss}
\end{minipage}
\hfill
\begin{minipage}[t]{0.45\linewidth}
\centering
\includegraphics[width=70mm]{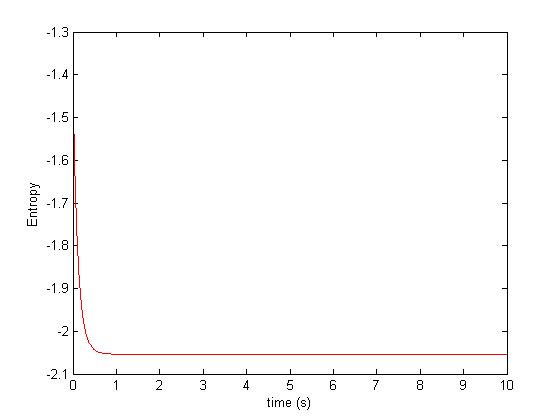}
\caption{The Entropy decay of numerical solution}\label{Entropy_FPL_2Gauss}
\end{minipage}
\end{figure}
The entropy decays to its equilibrium state fast and keeps stable after that. The whole decay process preserves mass, momentum and kinetic energy.
See Figure \ref{Evo_Pdf_FPL_2Gauss} and \ref{Entropy_FPL_2Gauss}.

\subsubsection{Multi-component Plasmas}\label{sec:numApp_2compPlasma}

In this section, we apply our scheme to a specific example of electro-neutral hydrogen plasma. The dimensionless system of equation writes \cite{Bobylev_MC_2plasma}
\begin{eqnarray}\label{Eqn_2plasma}
\frac{\partial f_{e}}{\partial t} &=& \frac{1}{2}\left[ Q^{(1)}_{FPL}(f_{e}, f_{e}) + Q^{(\theta)}_{FPL}(f_{e}, f_{i}) \right]  \nonumber \\
\frac{\partial f_{i}}{\partial t} &=& \frac{\theta^{2}}{2}\left[ Q^{(1)}_{FPL}(f_{i}, f_{i}) + Q^{(1/\theta)}_{FPL}(f_{i}, f_{e}) \right]\, ,
\end{eqnarray}
where $\theta\ll1$ is the dimensionless mass ratio of electrons to ions; the subscripts $e,i$ stand for electrons and ions respectively, with  
\begin{equation}\label{Q_theta}
Q^{(\theta)}_{FPL}(f,g)=\nabla_{v}\cdot\! \int \textbf{S}(v-v_{*})(f(v_{*})\nabla_{v}g(v)-\theta f(v)\nabla_{v_{*}}g(v_{*})) dv_{*}\, ,
\end{equation}
for any $\theta>0$, and the projection matrix $\textbf{S}$ as defined in (\ref{ProjMatrix}).
The corresponding time-dependent thermodynamic quantities associated to the electron/ion system are given by 
corresponding  time-dependent masses, mean velocities and temperatures for electrons and ions are defined by
\begin{align}\label{thermo-system}
\rho_{e}(t)=\frac{1}{3}\int f_{e}(v,t)dv,\ \  &\quad \ \ \rho_{i}(t)=\frac{1}{3\theta}\int f_{i}(v,t)dv\, ,\nonumber\\
\mu_{e}(t)=\frac{1}{3}\int f_{e}(v,t)v\,dv, &\quad \ \  \mu_{i}(t)=\frac{1}{3\theta}\int f_{i}(v,t)v\,dv\, .\\
T_{e}(t)=\frac{1}{3}\int f_{e}(v,t)|v|^{2}dv, &\quad  \ \ T_{i}(t)=\frac{1}{3\theta}\int f_{i}(v,t)|v|^{2}dv\, ,\nonumber
\end{align}
respectively.

This system of equations (\ref{Eqn_2plasma})  is endowed with normalized initial data as follows.   Their initial masses, $\rho_{e}(0)$ and $\rho_{i}(t)$ are normalized to unity.
Their means, $\mu_{e}(0)$ and $\mu_{i}(t)$ are null. Their  initial kinetic electron and ion  temperatures are finite numbers, that is 
$T_e(0)=T_{e,0}$ and $ T_i(0)=T_{i,0}$, respectively. This system conserves mass, momentum and energy according to the relations
\begin{align}\label{thermo-system-conserv}
\frac{\rho_{e}(t) +  \rho_{i}(t)}{2}  &\ =\  1 \quad ,  \ \ \ \ \mu_{e}(t) +  \mu_{i}(t) = 0 \qquad\ \  \text{and}\nonumber\\
T_{e}(t) + T_{i}(t) &\ =  \ \bar{T}\ :=\  T_{e,0} + T_{i,0},
\end{align}
respectively.

The corresponding  weak form of $Q^{(\theta)}_{FPL}(f,g)$ is given by
\begin{align}\label{weak-fg-form}
&\int Q^{(\theta)}_{FPL}(f,g)\varphi(v)dv \\
&=\iint f(v)g(v_{*})\left[ (1+\theta)\nabla_{v}\cdot\textbf{S}(v-v_{*}) \nabla\varphi(v) + \textbf{S}(v-v_{*}):\nabla^{2}\varphi(v) \right]dv_{*} dv\, ,\nonumber
\end{align}
where the Frobenius inner product $A:B=\text{Trace}(A^{T}B)$.

By taking the Fourier multiplier $\varphi(v)=(2\pi)^{-3/2}e^{-i\xi\cdot v}$,  as a test function in   \eqref{weak-fg-form}, yields a similar derivation, done as in (\ref{FPL_FT}),  in the spectral
representation of $Q$, 
\begin{equation}
\widehat{Q^{(\theta)}_{FPL}}(\widehat{f}, \widehat{g})(\xi)= \int \widehat{f}(\xi-w)\widehat{g}(w)\left[ (1+\theta)\xi^{T}\widehat{\mathbf{S}}(w)w - \xi^{T} \widehat{\mathbf{S}}(w) \xi\right]dw\, .
\end{equation}
\textit{Remark.} When $\theta=1$ and $f=g$ in (\ref{Q_theta}), the monoatomic case (\ref{Qhat_analytical}) is recovered.

We use the simulation of longtime dynamics associated to system  \eqref{Eqn_2plasma} as a validation and verification of our collisional spectral code with the constrain $L^2$ optimization algorithm that satisfies the conservation properties \eqref{thermo-system-conserv} associated to the system.

For such tests, we follow an analogous system to the one of A.V. Bobylev et al \cite{Bobylev_relax2plasma} performed for the radial (one-dimensional) Landau equations, now extended by our computational approach to non-isotropic distribution functions in 3-d velocity space.

In particular, we want to check that the electron and ion long time asymptotic temperatures satisfy the following system of ordinary differential equations, whose stationary states are exactly solvable as
well as stable. Under the assumption that $\theta\ll1$ and that initially $\theta T_{i,0} <1 T_{e,0} $, the set of ODEs governing the
relaxation of the two-temperature plasma is given by
\begin{eqnarray}\label{temp-system}
(\theta \bar{T}+(1-\theta)T_{e})^{\frac{3}{2}}\frac{dT_{e}}{dt}&=&\frac{4}{3\sqrt{2\pi}}(T_{e}-T_{i})\theta \, ,\\
T_{e}(t)+T_{i}(t)&=&\bar{T} \nonumber  \, ,
\end{eqnarray}
and the temperature difference follows
\begin{equation}\label{temp-diff-eq}
\frac{d(T_{i}-T_{e})}{dt}=-\theta\frac{8}{3\sqrt{2\pi}}\frac{T_{i}-T_{e}}{(\theta \bar{T}+(1-\theta)T_{e})^{\frac{3}{2}}}\,  \ ,
\end{equation}
which implies
\begin{equation}\label{temp-infty}
|T_{i}-T_{e}|\rightarrow 0, \quad \text{as }t\rightarrow\infty \, .
\end{equation}

So, when $t$ is large enough, or when the system approaches equilibrium, $T_{e}\approx T_{i}\approx\frac{\bar{T}}{2}$, the difference of temperatures
decays ``almost" exponentially (note that this is an approximation)
\begin{equation}\label{eqn:expdecay_2plasma}
|T_{i}(t)-T_{e}(t)|\approx |T_{i,0}-T_{e,0}|\text{exp}\left( -\frac{16}{3\sqrt{\pi}}\frac{\theta}{\left((1+\theta)\bar{T}\right)^{3/2}}t\right)\, .
\end{equation}

\begin{figure}[!htb]
\begin{minipage}[t]{0.45\linewidth}
\centering
\includegraphics[width=70mm]{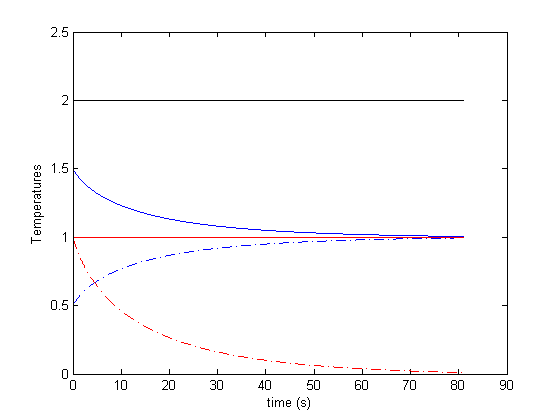}
\caption{The relaxation of temperatures for the 2-plasma system \eqref{Eqn_2plasma}: solid blue line: temperatures of ions; dash-dot blue: temperatures of electrons; top solid black: the total temperature (as defined in identity\eqref{temp-system}; bottom dash-dot red: temperature difference \eqref{temp-infty}   }\label{Temperature_2plasma}
\end{minipage}
\hfill
\begin{minipage}[t]{0.45\linewidth}
\centering
\includegraphics[width=70mm]{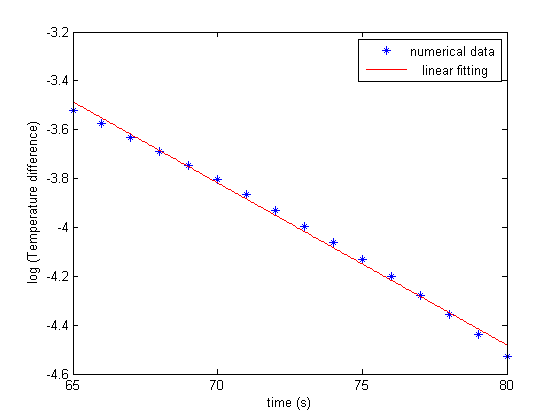}
\caption{The logarithm of temperature difference for large time and its linear fitting}\label{RelaxationRate_2plasma}
\end{minipage}
\end{figure}

We solve the equation system (\ref{Eqn_2plasma}) by our conservative scheme introduced above and observe the relaxation of temperatures for electrons and ions.
The dimensionless mass ratio used is $\theta=\frac{1}{16}$. The initial states are two Maxwellians for hot ions and cold electrons, say $T_{e}=\frac{1}{2}$ and $T_{i}=\frac{3}{2}$ (then $\bar{T}=2$) in (\ref{thermo-system-conserv}).
Figure \ref{Temperature_2plasma} shows the decay to equilibrium of the 2-plasma system as expected. If we take the logarithm of the temperature difference in (\ref{eqn:expdecay_2plasma}), we
can actually expect to observe the exponential decay rate in (\ref{eqn:expdecay_2plasma}), which is $-\frac{16}{3\sqrt{\pi}}\frac{\theta}{\left((1+\theta)\bar{T}\right)^{3/2}}=-0.061$ in this example.
Figure \ref{RelaxationRate_2plasma} shows the logarithm of the temperature difference (scattered data) when time is large enough (states approaching equilibrium) and its linear fitting, with a slope of -0.066343, which is a rough verification of our analytical prediction.

\subsection{Electron Plasma Waves}
 In this and following sections, our target is a two-species plasma system of electrons and ions. 
 
 In most plasma of interest, the ion temperature is much smaller than the electron temperature. Together with the fact that electrons have much smaller mass, the ions may be assumed to be stationary. If we assume the temperature of the ions is negligible compared to that of electrons, i.e $T_{i}/T_{e}\sim 0$, we may assume the ions obey a Dirac measure \cite{FPL_1x3v} (see \cite{Modelcollisions} for physical derivations),
 \begin{equation}
 f_{i}(t,x,v)=\rho_{i}(t,x)\delta_{0}(v-\bar{v}_{i})\, ,
 \end{equation}
 where the ion density $\rho_{i}$ and mean velocity $\bar{v}_{i}$ are given or satisfy certain hydrodynamic equations. Then, we get the ion-electron collision operator
 \begin{equation}\label{eqn:ele-plasma_linearQ}
 Q_{e,i}(f_{e})=\rho_{i}\nabla_{v}\cdot(\mathbf{S}(v-\bar{v}_{i})\nabla_{v}f_{i}(v))\, ,
 \end{equation}
 which is basically a linear operator w.r.t distribution $f_{i}$.

 The weak form of (\ref{eqn:ele-plasma_linearQ}) reads
 \begin{equation}
 \int_{\mathbb{R}^{3}}Q_{e,i}(f_{e})\varphi(v)dv=-\rho_{i}\int_{\mathbb{R}^{3}}(\mathbf{S}(v-\bar{v}_{i})\nabla_{v}f_{i}(v))\cdot \nabla_{v}\varphi(v)dv\, ,
 \end{equation}
 from which it is not difficult to prove that the linear operator (\ref{eqn:ele-plasma_linearQ}) conserves mass and energy, by noticing that the zero eigen-space of projection matrix $\mathbf{S}(v)$ is spanned by $v$.

 Similar to the spectral representation of the fully nonlinear collision operator (\ref{FPL_FT}), we can also obtain the spectral representation for (\ref{eqn:ele-plasma_linearQ})
 \begin{equation}
 \begin{split}
 \widehat{Q}_{e,i}(\widehat{f}_{e}) &= i(2\pi)^{-3/2}\int_{\mathbb{R}^{3}}\xi^{T}\mathbf{S}(v)\nabla_{v}f_{e}\exp(-iv\cdot\xi)dv \\
 &=-(2\pi)^{-3/2}\int_{\mathbb{R}^{3}}\xi^{T}\mathbf{S}(w)(\xi-w)\widehat{f}_{e}(\xi-w)dw\, .
 \end{split}
 \end{equation}

 Since the conservation routine (see Section \ref{sec:linking}) can force the conservation of any desired moments, we have to adjust it
 for the linear operator (\ref{eqn:ele-plasma_linearQ}), which only conserves mass and energy. This is done by choosing a new $2\times 2N^{3}$ constraint matrix by extracting only the first and fifth (in 3D case) rows of the full $5\times 2N^{3}$ constraint matrix (\ref{computeC2}).

 Then, the final model for electron plasma waves reads
 \begin{equation}
 \label{eqn:plasmaWaves}
 \frac{\partial}{\partial t}f_{e}+v\cdot \nabla_{x}f_{e}+E(t,x)\cdot\nabla_{v}f_{e} = \frac{1}{\varepsilon}\big(Q_{e,e}(f_{e},f_{e}) + Q_{e,i}(f_{e}) \big)\, ,
 \end{equation}
 which will be solved by the combined RKDG-Spectral method developed in this article.

 \subsection{The Linear Landau Damping}

Perhaps, one of the most astonishing theoretical discoveries of plasma physics is the wave damping without energy dissipation by collisions. It is a result of wave-particle interactions. It occurs due to the energy exchange between particles in motion in the plasma and an electromagnetic wave. The velocity of a particle may be greater or less than the phase velocity of the wave. Thus, there are particles gaining energy from the wave and leading to wave damping, and also, there are particles losing energy to the wave and resulting in a increase of the wave energy. The Landau damping is studied by perturbing the Maxwellian distribution by a wave. An extremely small wave amplitude will restrict the problem in a linear regime, and thus lead to problem of ``linear Landau damping". However, if the wave amplitude is relatively large, we are in a regime of ``nonlinear Landau damping". In this section, we study the linear damping first.

The initial condition is taken as a small of perturbation of the global equilibrium $M(v)=(2\pi)^{-\frac{3}{2}}\exp(-\frac{|v|^{2}}{2})$
\begin{equation}\label{eqn:init_lineardamping}
f_{0}(x,v)=(1+A\cos(kx)) M(v)\, ,
\end{equation}
for $(x,v)\in[0,2\pi/k] \times \mathbb{R}^{3}$. Such an initial state has been chosen by many authors, see for instance \cite{FPL_1x3v,CGM_RKDG_VP}, as a benchmark problem for studying damping properties.

To study linear damping, we have to make the amplitude small enough, e.g. $A=10^{-5}$, to restrict the problem under linear regimes. To well capture the Landau damping, the velocity domain must be large enough. It has to be larger than the phase velocity $v_{\phi}=\omega/k$, where $\omega$ is the frequency approximated by \cite{FPL_1x3v}
\begin{equation}
 \omega^{2}=1+3k^{2}\, .
\end{equation}
Here, we select $L_{v}=5.75$.

The classical Landau theory tells that the square root of the electrostatic energy
\begin{equation}
 \frac{1}{2}\int^{L_{x}}_{0} |E_{h}(x)|^{2} dx
\end{equation}
is expected to decay exponentially with frequency $\omega$. We will plot the evolution of logarithm of square root of the electrostatic energy and compute its numerical damping rate.

According to \cite{Chen_PlasmaPhysics,Delcroix_plasma}, the theoretical damping rate can be estimated as
\begin{equation}\label{eqn:dampingrate}
\lambda = \lambda_{l} + \lambda_{c}\, ,
\end{equation}
where $\lambda_{l}$ is the damping rate for collisionless plasma and $\lambda_{c}$ is the ``correction" for collisional case.
\begin{equation}
\lambda_{c} = -\frac{\nu}{3}\sqrt{\frac{2}{\pi}}\, ,
\end{equation}
with $\nu=\frac{1}{\varepsilon}$ denoting the collision frequency. And, $\lambda_{l}$ is estimated by
\begin{equation} \label{eqn:dampingrate0}
\lambda_{l} = -\sqrt{\frac{\pi}{8}}\frac{1}{k^{3}}\exp(-\frac{1}{2k^{2}}-\frac{3}{2})\, .
\end{equation}
However, as pointed out in \cite{FPL_1x3v}, (\ref{eqn:dampingrate0}) is more accurate when wave number $k$ is large; so, for small wave numbers, more accurate estimate is available in \cite{Accuratedamping}
\begin{equation}\label{eqn:dampingrate1}
\lambda_{l} =-\sqrt{\frac{\pi}{8}}\left(\frac{1}{k^{3}}-6k \right)\exp(-\frac{1}{2k^{2}}-\frac{3}{2}-3k^{2}-12k^{4})\, ,
\end{equation}
and frequency
\begin{equation}
\omega = 1 + 3k^{2}+6k^{4}+12k^{6}\, .
\end{equation}

We will test with initials (\ref{eqn:init_lineardamping}) for both collisionless and collisional cases.

\begin{figure}[!htb]\label{fig:damping_k05}
\begin{minipage}[t]{0.45\linewidth}
\centering
\includegraphics[width=75mm]{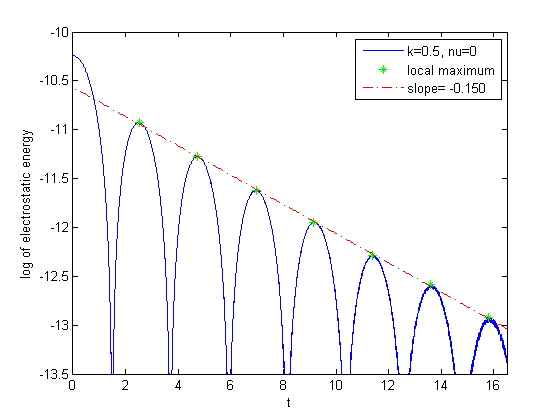}
\end{minipage}
\hfill
\begin{minipage}[t]{0.45\linewidth}
\centering
\includegraphics[width=75mm]{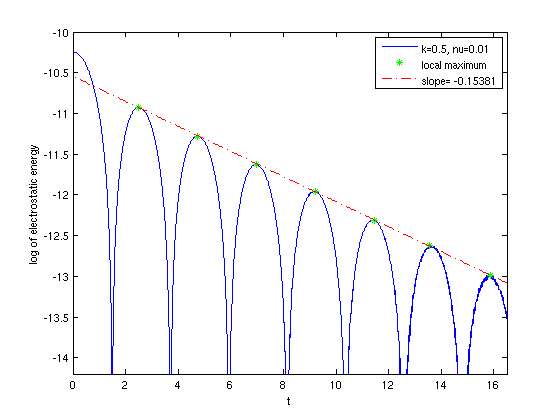}
\end{minipage}
\caption{Linear Landau damping for wave number $k=0.5$: $\varepsilon=\infty$ (left), $\varepsilon=100$ (right)}
\end{figure}

We assume $\rho_{i}=1$ and $\bar{v}_{i}=0$, and fix wave number $k=0.3, 0.5$. Since here amplitude $A$ is small enough, the model can be seen in its linear regime and we can compare the numerical damping results against theoretical predictions (\ref{eqn:dampingrate}).
Our numerical results recovered the exponential damping behaviors and show that the damping is stronger if collisions are taking effects. And the damping rate increases with larger wave number $k$.
In collisionless case, i.e $\varepsilon=\infty$, when $k=0.5$, formula (\ref{eqn:dampingrate0}) gives an estimation $-0.151$ which agrees well with our numerical result in Figure \ref{fig:damping_k05}; but for $k=0.3$, formula (\ref{eqn:dampingrate1}) gives a more accurate estimate $-0.0132$ (formula (\ref{eqn:dampingrate0}) gives $-0.020$). In collisional case, e.g $\varepsilon=100$, theoretically estimated damping rate for $k=0.5$ is $-0.154$, while for $k=0.3$ is $-0.0167$.
Also, from the damping result, we know larger collision frequency impose a stronger damping.

\begin{figure}[!htb]\label{fig:damping_k03}
\begin{minipage}[t]{0.45\linewidth}
\centering
\includegraphics[width=75mm]{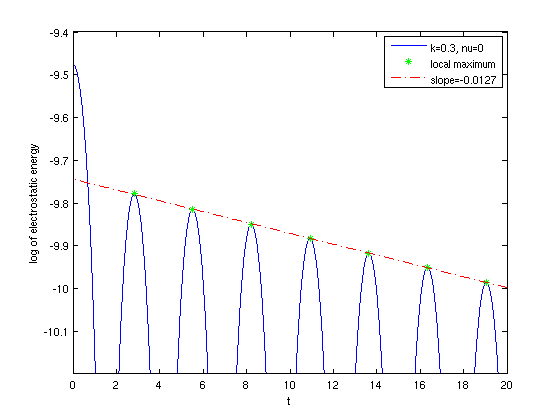}
\end{minipage}
\hfill
\begin{minipage}[t]{0.45\linewidth}
\centering
\includegraphics[width=75mm]{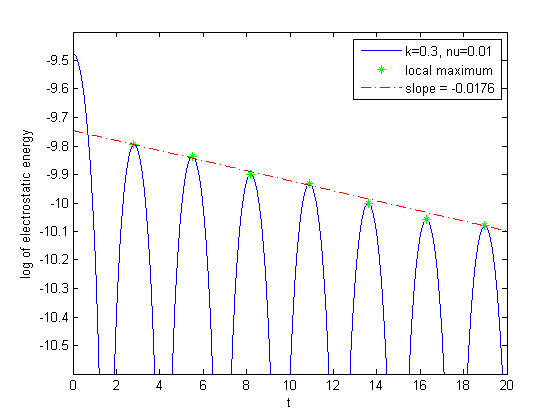}
\end{minipage}
\caption{Linear Landau damping for wave number $k=0.3$: $\varepsilon=\infty$ (left), $\varepsilon=100$ (right)}
\end{figure}

\subsection{The Nonlinear Landau Damping}

The linear theory regarding plasmas has been relatively well developed (though still many problems remain unsolved). However, the nonlinear phenomena of plasma is much less understood. From last section, we know as long as the wave amplitude $A$ is small enough, a well-developed linear
 theory is valid. Nevertheless, when the wave amplitude gets larger, many waves in experiments can no longer be described by the linear theory. Some of them are not even trackable through analysis.

 One example would be ``electron trapping" phenomenon, which occurs with the nonlinear Landau damping of the waves. Since the particles travel relative to the wave, a large electric potential together with collisions will trap the electrons in a potential well of the wave. The trapped electrons will be bounced back and forth in the well, causing fluctuating amplitudes of the wave. Thus, one cannot always expect an exponential damping as in the linear case.

 In order to capture the electron trapping, we extract the contours of the following marginal distribution
 \begin{equation}
 F(t,x, v_{x})=\int_{\mathbb{R}^{2}}f(t,x,v_{x},v_{y},v_{z})dv_{y}dv_{z} \, .
 \end{equation}

 In phase space, $F(t,x,v_{x})$ will form peaks whenever there is a potential trough. Trapped electrons will move in closed orbits in phase space, since the contours $F(t,x,v_{x})$ are also the electron trajectories. Please refer to \cite{Chen_PlasmaPhysics} for more explanations.

 In this section, we will study the nonlinear damping with the following initial wave
\begin{equation}\label{eqn:init_nonlineardamping}
f_{0}(x,v)=(1+A\cos(kx)) M(v)\, , (x,v)\in[0,2\pi/k] \times \mathbb{R}^{3}\, ,
\end{equation}
for a relatively large amplitude $A$ such that it is no longer in the linear regime. Here, we choose the Maxwellian
\begin{equation*}
M(v)=(2\pi T)^{-\frac{3}{2}}\exp(-\frac{|v|^{2}}{2T})\, .
\end{equation*}

\begin{figure}[!htb]
\centering
\begin{minipage}{0.32\linewidth}
\centering
\includegraphics[width=\linewidth]{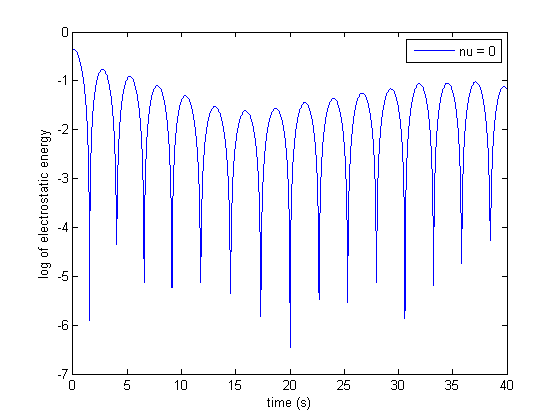}
\end{minipage}\hfill
\begin{minipage}{0.32\linewidth}
\centering
\includegraphics[width=\linewidth]{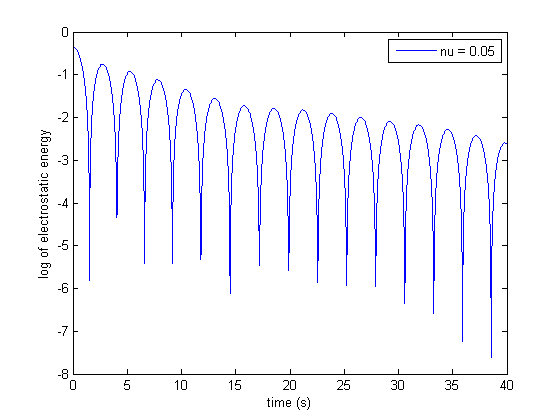}
\end{minipage}\hfill
\begin{minipage}{0.32\linewidth}
\centering
\includegraphics[width=\linewidth]{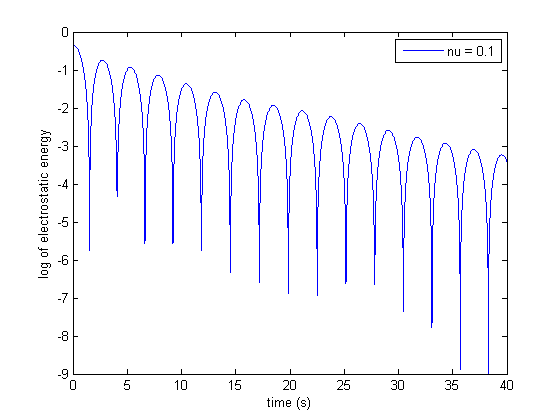}
\end{minipage}
\caption{Nonlinear damping with $A=0.2$ for $\nu=0$ (left), $\nu=0.05$ (middle) and $\nu=0.1$ (right)}\label{fig:nonlinDamping}

\end{figure}

\begin{figure}[!htb]
\centering
\includegraphics[width=90mm]{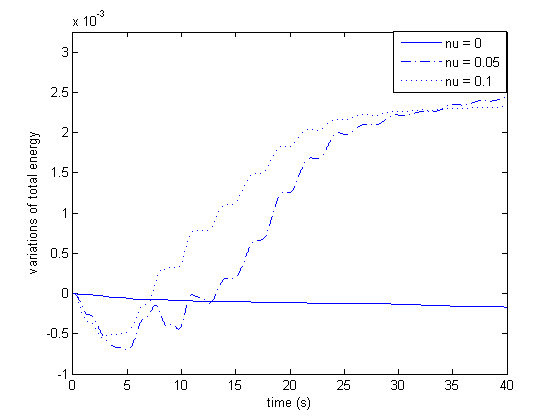}
\caption{Relative errors of total energy $e_{tot}$ {(as defined in identity \eqref{eqn:totalEne})}  during nonlinear damping simulation with $A=0.2$ for $\nu=0,0.05,0.1$}\label{fig:VarTotEne_nonlinDamping}
\end{figure}

Figure \ref{fig:nonlinDamping} shows the nonlinear damping results for $A=0.2$, $T=0.5$, $k=0.5$ and a large enough velocity domain $L_{v}=5$, with different collision frequencies $\nu=0,0.05,0.1$. We choose $N_{x}=36$ mesh elements in $x$-direction, $N_{v}=36$ mesh elements in each direction of velocity $v$ for the RKDG VP problem, and $N=24$ Fourier modes for the spectral method. We can see the electric energy, in all cases, decreases exponentially at first. In the collisionless regime, the electric energy then starts to oscillate around a constant, which agrees well with the known property. With collisions, the oscillations are weakened. In particular, with the presence of stronger collisions, the amplitude of electric energy will start to form an exponential decay again. Although a relatively large amplitude $A$ is imposed and moderate resolution of mesh is applied, we still obtain a good preservation of the total energy, which is even better conserved than \cite{FPL_1x3v}. See Figure \ref{fig:VarTotEne_nonlinDamping} for variations of total energy during the whole process of simulation. Here the total energy being computed is the sum of kinetic energy plus electrostatic energy, i.e 
\begin{equation}\label{eqn:totalEne}
e_{tot} = \frac{1}{2}\int^{L_{x}}_{0}\int_{\mathbb{R}^{3}}f(x,v,t)|v|^{2}dv dx + \frac{1}{2}\int^{L_{x}}_{0} |E_{h}(x)|^{2} dx
\end{equation}

\begin{figure}[!htb]

\centering

\begin{minipage}{0.32\linewidth}
\centering
\includegraphics[width=\linewidth]{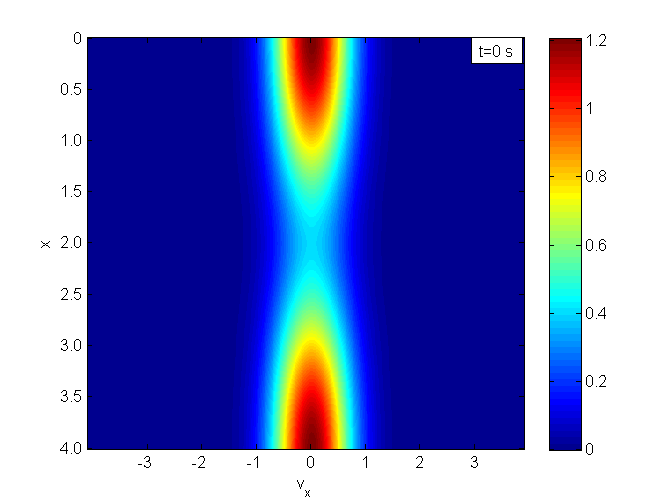}
\end{minipage}\hfill
\begin{minipage}{0.32\linewidth}
\centering
\includegraphics[width=\linewidth]{EleTrapping_smoothed_t0.png}
\end{minipage}\hfill
\begin{minipage}{0.32\linewidth}
\centering
\includegraphics[width=\linewidth]{EleTrapping_smoothed_t0.png}
\end{minipage}

\begin{minipage}{0.32\linewidth}
\centering
\includegraphics[width=\linewidth]{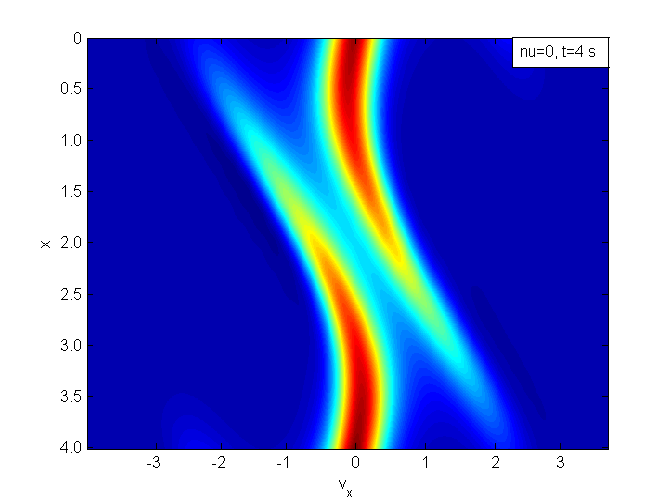}
\end{minipage}\hfill
\begin{minipage}{0.32\linewidth}
\centering
\includegraphics[width=\linewidth]{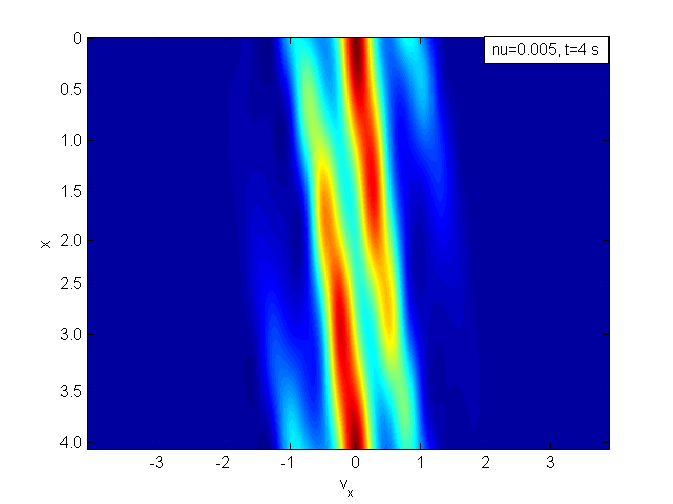}
\end{minipage}\hfill
\begin{minipage}{0.32\linewidth}
\centering
\includegraphics[width=\linewidth]{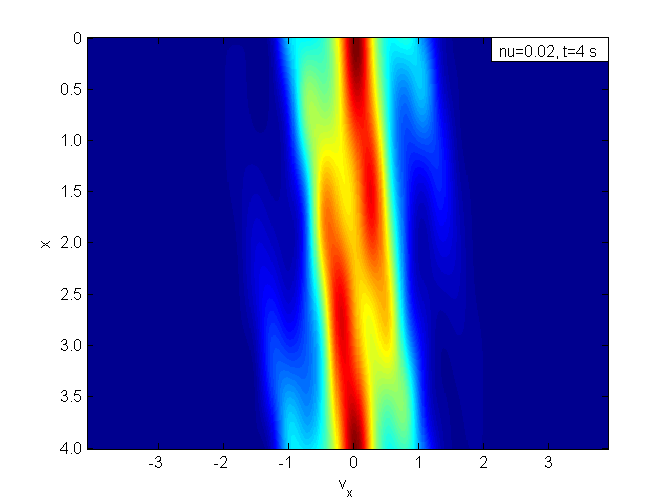}
\end{minipage}

\begin{minipage}{0.32\linewidth}
\centering
\includegraphics[width=\linewidth]{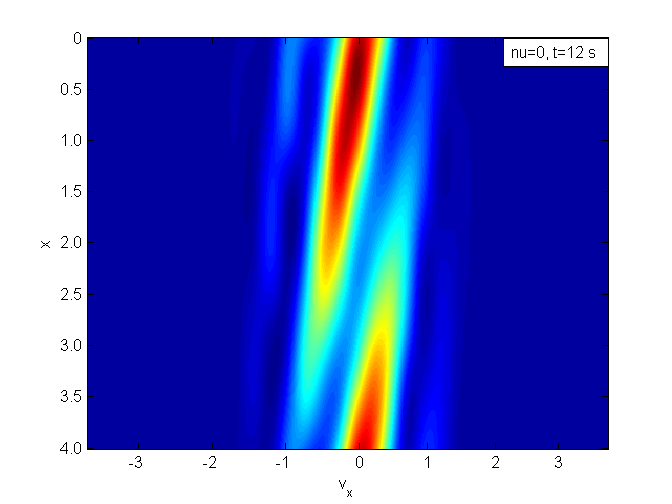}
\end{minipage}\hfill
\begin{minipage}{0.32\linewidth}
\centering
\includegraphics[width=\linewidth]{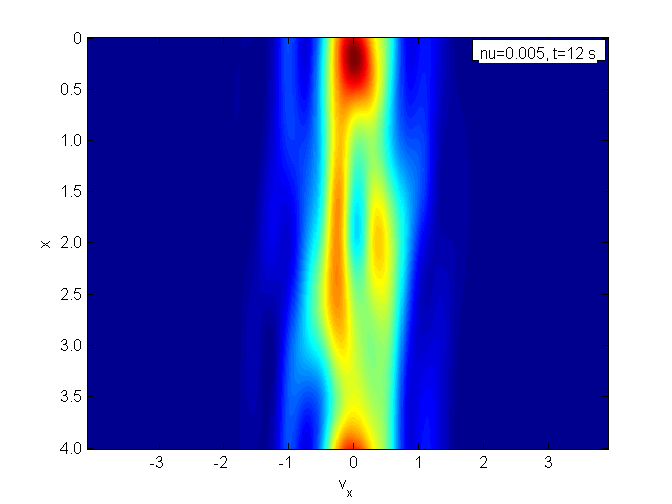}
\end{minipage}\hfill
\begin{minipage}{0.32\linewidth}
\centering
\includegraphics[width=\linewidth]{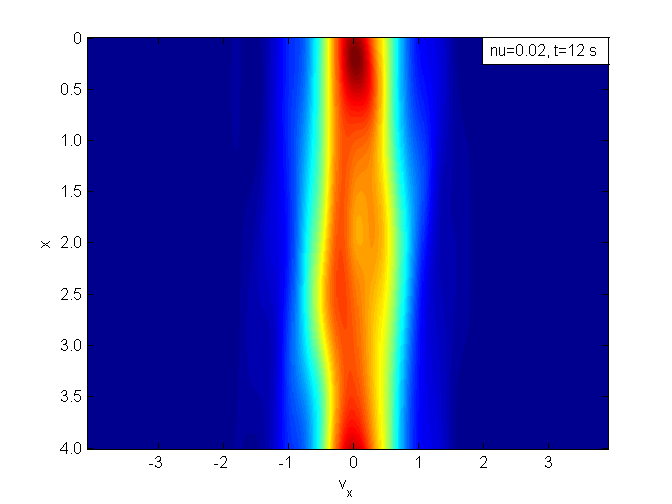}
\end{minipage}

\begin{minipage}{0.32\linewidth}
\centering
\includegraphics[width=\linewidth]{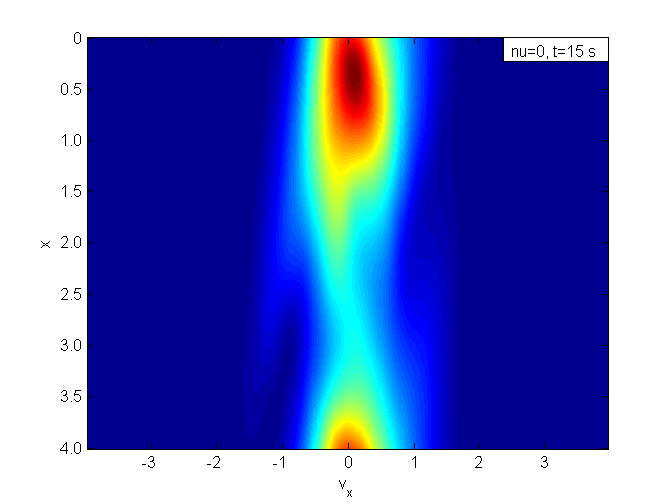}
\end{minipage}\hfill
\begin{minipage}{0.32\linewidth}
\centering
\includegraphics[width=\linewidth]{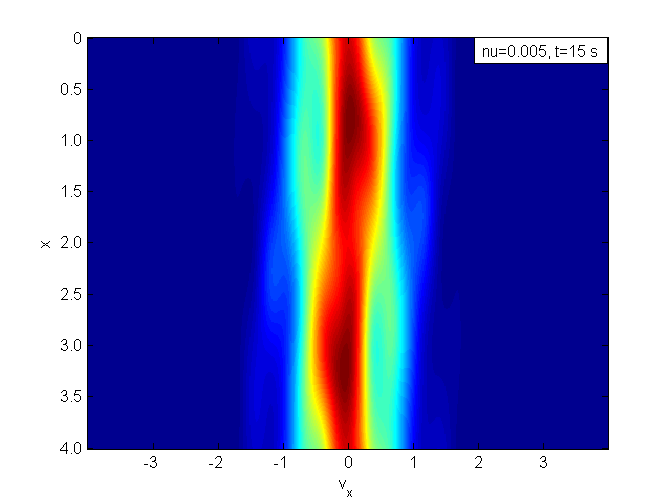}
\end{minipage}\hfill
\begin{minipage}{0.32\linewidth}
\centering
\includegraphics[width=\linewidth]{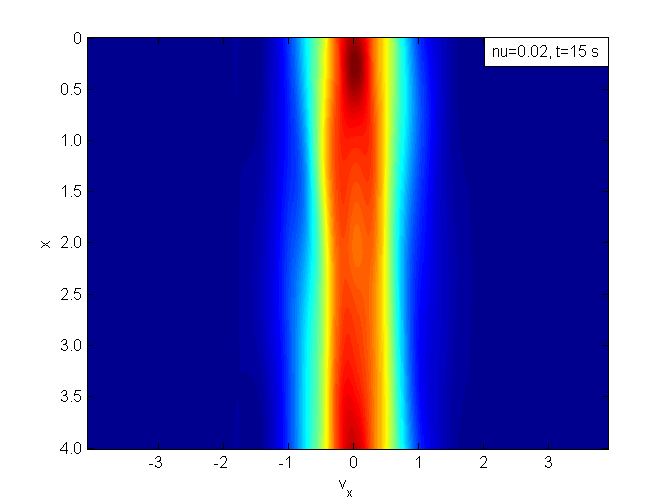}
\end{minipage}

\begin{minipage}{0.32\linewidth}
\centering
\includegraphics[width=\linewidth]{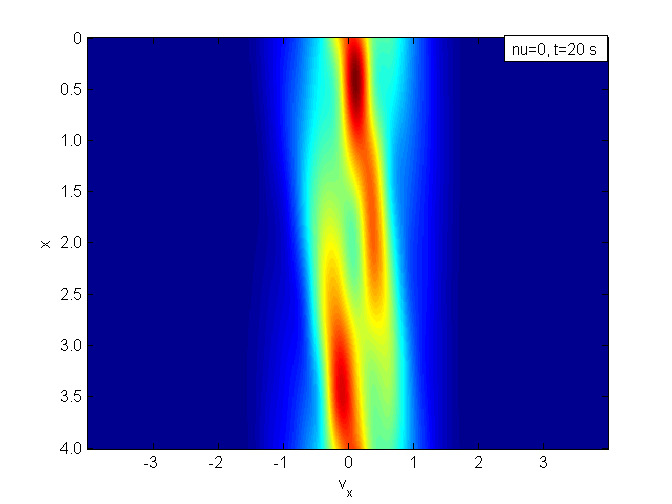}
\end{minipage}\hfill
\begin{minipage}{0.32\linewidth}
\centering
\includegraphics[width=\linewidth]{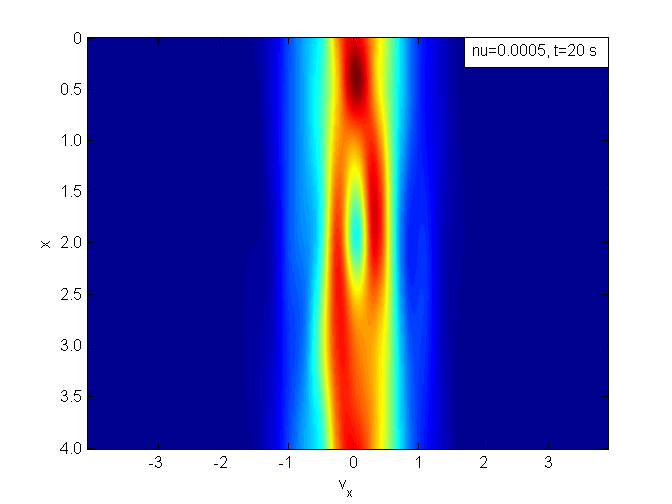}
\end{minipage}\hfill
\begin{minipage}{0.32\linewidth}
\centering
\includegraphics[width=\linewidth]{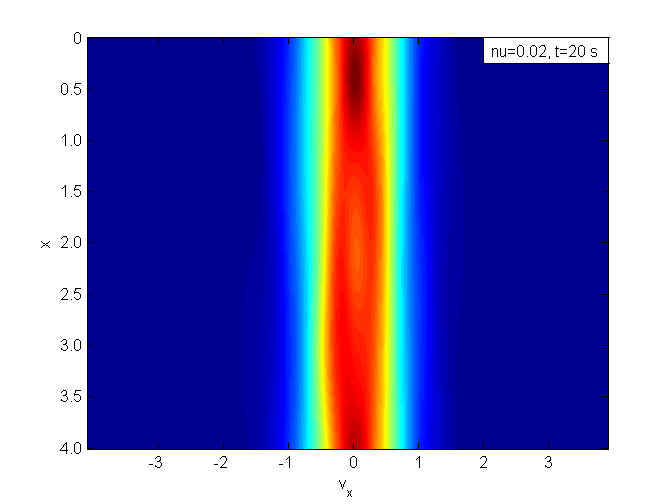}
\end{minipage}
\caption{Evolution of $F(t,x,v_{x})$ for $\nu=0$ (left), $\nu=0.005$ (middle) and $\nu=0.02$ (right)}\label{fig:eletrapping}
\end{figure}

Figure \ref{fig:eletrapping} shows the electron trapping effects for much larger amplitude $A=0.5$, $T=0.25$, $k=2\pi/4$ and $L_{v}=4$. We choose $N_{x}=48$, $N_{v}=32$ and $N=24$. Collision effects range from weak to strong , that is, $\nu=0,0.005,0.02$.  One can observe that, without collisions, more electrons are trapped in the potential trough. When collisions get stronger, less and less electrons are trapped and a stationary state is reached early.

\subsection{Two Stream Flow}
This is of primary importance for studying nonlinear effects of plasmas in future.
In this section, we consider a plasma with fixed ion background and only consider the electron-electron collisions. {This is basically a single-carrier problem modeled by (\ref{eqn:plasmaWaves}) without the electron-ion collision terms.} We will study how well the above time-splitting and conservative linking process work, by initializing with a non-isotropic two-stream flow.
\begin{equation}
 f_{0}(x,v)=(1+A\cos(kx))f_{TS}(v)\, ,
\end{equation}
where $A$ is the amplitude of the perturbation and $k$ the wave number, and
\begin{equation}
 f_{TS}(v)=\frac{1}{2(2\pi\sigma^{2})^{3/2}}\left[ \text{exp}\left(-\frac{|v-2\sigma e|^{2}}{2\sigma^{2}} \right) + \text{exp}\left(-\frac{|v+2\sigma e|^{2}}{2\sigma^{2}} \right) \right]\, ,
\end{equation}
with parameter $\sigma=\pi/10$ and $e=(1,0,0)$. We would like it to be far from the linear regime, so a relatively large perturbation is considered $A=0.5$, $k=2\pi/L_{x}$ with $L_{x}=4$.
A large enough velocity domain is selected $L_{v}=4.5$. We choose $N_{x}=48$ mesh elements in $x$-direction, $N_{v}=32$ mesh elements on each direction of velocity $v$ for the RKDG VP problem,
and $N=24$ Fourier nodes for the spectral method.

\begin{figure}[!htb]
\begin{minipage}[t]{0.45\linewidth}
\centering
\includegraphics[width=70mm]{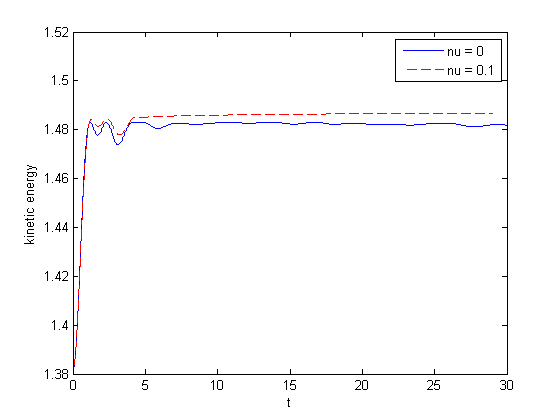}
\caption{The evolution of kinetic energy for the two-stream flow}\label{fig:inhomoLP_2stream_kineticEne}
\end{minipage}
\hfill
\begin{minipage}[t]{0.45\linewidth}
\centering
\includegraphics[width=70mm]{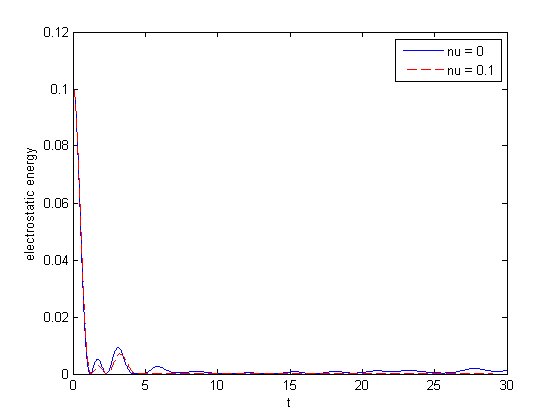}
\caption{The evolution of electrostatic energy for the two-stream flow}\label{fig:inhomoLP_2stream_elecEne}
\end{minipage}
\end{figure}

\begin{figure}[!htb]

\centering
\includegraphics[width=90mm]{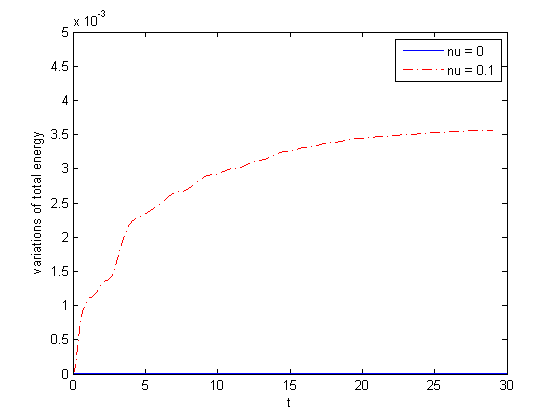}
\caption{Relative errors of total energy $e_{tot}$ {(as defined in identity \eqref{eqn:totalEne})} for the two-stream flow}\label{fig:inhomoLP_2stream_varEne}

\end{figure}

A quite strong collision effect is considered by taking a relatively large collision frequency $\nu=0.1$ (relatively small Kundsen number $\varepsilon=10$). Results are also compared to collisionless case, i.e. $\nu=0$. In Figure~(\ref{fig:inhomoLP_2stream_kineticEne}), (\ref{fig:inhomoLP_2stream_elecEne}) and (\ref{fig:inhomoLP_2stream_varEne}), the total energy $e_{tot}$, defined in \eqref{eqn:totalEne}, initially comes from both the kinetic and electrostatic energy, but with time forwarding, the electrostatic energy decays with oscillations down
to zero and the total energy at the end all comes from pure kinetic motions. This means the system has reached its global equilibrium. During the whole process, the total energy is well preserved only with negligible variations, even with quite strong collision $\nu=0.1$. In addition, from Figure (\ref{fig:inhomoLP_2stream_kineticEne}) and Figure (\ref{fig:inhomoLP_2stream_elecEne}) one can observe that, since the Landau operator is essentially a diffusive operator, the oscillations generated by coupling with the Poisson equations damps with collisions, and thus the state reaches stationary in a much earlier stage.

We finally note that the relative  error of total energy  computed, both for the non-linear electron Landau damping Figure~(\ref{fig:VarTotEne_nonlinDamping}), and for  the electron two-stream flow Figure~(\ref{fig:inhomoLP_2stream_varEne})
are bounded in time, uniformly in the relatively small Kundsen number chosen for these simulations.
While the error analysis of the complete numerical scheme has not been performed up to date,  such errors diminish as the number of mesh
points increases in the RKDG-VP approximation, as well as when number of Fourier modes increases in the spectral approximation of the FPL operator.
 
\section{Summary and Future Work}

We studied the inhomogeneous FPL equations, coupled with the Poisson equations governing the self-consistent electric field.
The complicated inhomogeneous problem was split into two subproblems, by time-splitting scheme. We applied two different methods for treating the pure transport Vlasov problem and the pure collisional
homogeneous FPL equation. The former was solved by RKDG method, which had achieved its success in many other kinetic problems; while the latter was treated using conservative spectral method.
The conservative spectral method was well developed for solving Boltzmann equations and we extended the method to the FPL problems and applied it to study the multi-component plasma. The temperature relaxation of the multi-component plasma was studied both
analytically and numerically.  To link the two different methods, or computing grids, we developed a new conservation routine which can guarantee no loss of moments when projecting the Fourier solution onto
DG meshes. All desired moments are preserved only with error of DG approximations. The whole scheme has been applied to study the well-known Landau damping problems, whose results agree well with theoretical estimates,
and to two stream flows.

The project was implemented with parallelization, hybrid MPI \cite{MPI} and OpenMP \cite{Openmp}.

In the future, we would like to speed up the collision and conservation processes and increase the grid resolution, such that the current solver can tackle, in real world of collisional plasma, more challenging problems which is tough to be treated numerically. For example, we would like to apply non-periodic boundary conditons on the Poisson equation and thus to study more nonlinear effects, for instance plasma sheath problems, which is of
primary importance for Aerospace Engineering. 

\section*{Acknowledgement}
The authors thank Jeffrey Haack for very valuable conversations and Clark Pennie for many suggestions that improved this manuscript presentation.
The work of both authors has been partially supported by the NSF under grants DMS-1413064, DMS-1217154 and NSF-RNMS 1107465. Support from  the Institute of Computational Engineering and Sciences (ICES) at the University of Texas Austin is gratefully acknowledged.





\section*{Appendix -- Calculations of $\widehat{\mathbf{S}}$}

\noindent (1). $\widehat{\textbf{S}^{1}_{11}}(\omega)$.

This is done immediately.
\begin{equation}\label{S1hat}
\begin{split}
\widehat{\textbf{S}^{1}}(\omega)&=(2\pi)^{-3/2}\int_{B_{R}(0)}\frac{1}{|u|}e^{-i\omega\cdot u}du \\
&=\sqrt{\frac{2}{\pi}}\frac{1}{|\omega|^{2}}[1-\cos(R|\omega|)]
\end{split}
\end{equation}
And, if $|\omega|=0$, $\widehat{\textbf{S}^{1}}(\omega)=\sqrt{\frac{1}{2\pi}}R^{2}$.

\noindent (2). $\widehat{\textbf{S}^{2}_{33}}(\omega)$.

\begin{equation}\label{S233hat}
\begin{split}
\widehat{\textbf{S}^{2}_{33}}(\omega) &=(2\pi)^{-3/2}\int_{B_{R}(0)}\frac{u^{2}_{3}}{|u|^{3}}e^{-i\omega\cdot u}du \\
&=(2\pi)^{-3/2} \int^{R}_{0}r \int_{S^{2}} \sigma^{2}_{3}e^{-i r\omega\cdot \sigma}d\sigma dr
\end{split}
\end{equation}

Suppose $\omega=|\omega|(\sin(\theta)\cos(\phi),
\sin(\theta)\sin(\phi),\cos(\theta))^{T}$, and consider the
orthogonal rotation matrices
\begin{equation}\label{RotMat}
    R_{y}(\theta)=\left(
                 \begin{array}{ccc}
                   \cos(\theta) & 0 & \sin(\theta) \\
                    0 & 1 & 0 \\
                    -\sin(\theta) & 0 & \cos(\theta) \\
                 \end{array}
               \right)
\quad \text{and}\quad
    R_{z}(\phi)=\left(
                 \begin{array}{ccc}
                   \cos(\phi) &  \sin(\phi) & 0 \\
                    -\sin(\phi) & \cos(\phi) & 0\\
                     0 & 0 & 1 \\
                 \end{array}
               \right)
\end{equation}
which rotates the vectors about $y-$ and $z-$axis, respectively.

Then,
\begin{equation*}
R^{T}_{y}(\theta)R_{z}(\phi)\omega=\left( 0, 0, |\omega| \right)^{T} :=\tilde{\omega}
\end{equation*}
Denote $A=R^{T}_{y}(\theta)R_{z}(\phi)$, then A is also an orthogonal rotation matrix
\begin{equation}\label{RotMat}
    A=\frac{1}{|\omega|}\left(
                 \begin{array}{ccc}
                   \frac{\omega_{1}\omega_{3}}{\sqrt{\omega^{2}_{1}+\omega^{2}_{2}}} &  \frac{\omega_{2}\omega_{3}}{\sqrt{\omega^{2}_{1}+\omega^{2}_{2}}} & -\sqrt{\omega^{2}_{1}+\omega^{2}_{2}} \\
                    -\frac{\omega_{2}|\omega|}{\sqrt{\omega^{2}_{1}+\omega^{2}_{2}}} & \frac{\omega_{1}|\omega|}{\sqrt{\omega^{2}_{1}+\omega^{2}_{2}}}  & 0\\
                     \omega_{1} & \omega_{2} & \omega{3} \\
                 \end{array}
               \right)
\end{equation}
where we assume $\omega^{2}_{1}+\omega^{2}_{2} \neq 0$; otherwise, matrix $A$ is reduced to the identity matrix.

Then
\begin{equation*}
\begin{split}
 & \int_{S^{2}} \sigma^{2}_{3}e^{-i r \omega\cdot \sigma}d\sigma \\
&=\frac{1}{|\omega|^{2}} \big(4\pi(\omega^{2}_{1}+\omega^{2}_{2})\frac{\sin(r|\omega|)-r|\omega|\cos(r|\omega|)}{(r|\omega|)^{3}} \\
&\quad +4\pi\omega^{2}_{3}\frac{((r|\omega|)^{2}-2)\sin(r|\omega|)+2r|\omega|\cos(r|\omega|)}{(r|\omega|)^{3}}\big)
\end{split}
\end{equation*}

So, plugging back into $\widehat{\textbf{S}^{2}_{33}}$ \eqref{S233hat} gives
\begin{equation}
\begin{split}
\widehat{\textbf{S}^{2}_{33}}(\omega) &=(2\pi)^{-3/2} \int^{R}_{0}r \int_{S^{2}} \sigma^{2}_{3}e^{-i r\omega\cdot \sigma}d\sigma dr \\
&=\sqrt{\frac{2}{\pi}}\frac{1}{|\omega|^{4}}\big((\omega^{2}_{1}+\omega^{2}_{2})\frac{R|\omega|-\sin(R|\omega|)}{R|\omega|} \\
&\quad -\omega^{2}_{3}\frac{R|\omega| +R|\omega|\cos(R|\omega|)-2\sin(R|\omega|)}{R|\omega|}\big)
\end{split}
\end{equation}
And, if $|\omega|=0$, $\widehat{\textbf{S}^{2}_{33}}(\omega)=\sqrt{\frac{1}{2\pi}}\frac{R^{2}}{3}$.

\noindent (3). $\widehat{\textbf{S}^{2}_{13}}(\omega)$.

\begin{equation}\label{S213hat1}
\begin{split}
\widehat{\textbf{S}^{2}_{13}}(\omega) &=(2\pi)^{-3/2}\int_{B_{R}(0)}\frac{u_{1}u_{3}}{|u|^{3}}e^{-i\omega\cdot u}du \\
&=(2\pi)^{-3/2} \int^{R}_{0}r \int_{S^{2}} \sigma_{1}\sigma_{3}e^{-i r\omega\cdot \sigma}d\sigma dr
\end{split}
\end{equation}
Following the same change of variables as above,
\begin{equation*}
\begin{split}
 &\int_{S^{2}} \sigma_{1}\sigma_{3}e^{-i r\omega\cdot \sigma}d\sigma dr = \int_{S^{2}}(A^{T}\sigma)_{1}(A^{T}\sigma)_{3}e^{-ir \tilde{\omega}\cdot \sigma}d\sigma \\
 &=4\pi\frac{\omega_{1}\omega_{3}}{|\omega|^{2}}\frac{((r|\omega|)^{2}-3)\sin(r|\omega|)+3r|\omega|\cos(r|\omega|)}{(r|\omega|)^{3}}
\end{split}
\end{equation*}

So,
\begin{equation}\label{S213hat2}
\begin{split}
\widehat{\textbf{S}^{2}_{13}}(\omega) &=(2\pi)^{-3/2} \int^{R}_{0}r \int_{S^{2}} \sigma_{1}\sigma_{3}e^{-i r\omega\cdot \sigma}d\sigma dr \\
&=-\sqrt{\frac{2}{\pi}}\frac{\omega_{1}\omega_{3}}{|\omega|^{4}}\frac{2R|\omega| +R|\omega|\cos(R|\omega|) - 3\sin(R|\omega|)}{R|\omega|}
\end{split}
\end{equation}
And, if $|\omega|=0$, $\widehat{\textbf{S}^{2}_{13}}(\omega)=0$.

\end{document}